\documentclass[longauth]{aa}  

\usepackage{graphicx}
\usepackage{multirow}
\usepackage[dvipsnames]{xcolor}
\usepackage{placeins}
\usepackage{hyperref}
\usepackage{booktabs}
\usepackage{bm}
\hypersetup{colorlinks=true,citecolor=blue}

\newcommand{\tess}{\ensuremath{\emph{TESS}}\xspace}
\newcommand{\hst}{\ensuremath{\emph{HST}}\xspace}
\newcommand{\jwst}{\ensuremath{\emph{JWST}}\xspace}

\newcommand{\kepler}{\ensuremath{\emph{Kepler}}\xspace}
\newcommand{\sherlock}{\texttt{SHERLOCK}\,}
\newcommand{\tpfplotter}{\texttt{tpfplotter}\,}
\newcommand{\lightkurve}{\texttt{lightkurve}\,}
\newcommand{\triceratops}{\texttt{TRICERATOPS}\,}
\newcommand{\wotan}{\texttt{W{\={o}}tanS}\,}
\newcommand{\transitleastsquares}{\texttt{Transit Least Squares}\,}
\newcommand{\allesfitter}{\texttt{Allesfitter}\,}

\usepackage{txfonts}
\usepackage{tikz,xcolor,hyperref}
\definecolor{lime}{HTML}{A6CE39}
\DeclareRobustCommand{\orcidicon}{%
	\hspace{-1.5mm}
	\begin{tikzpicture}
	\draw[lime, fill=lime] (0,0) 
	circle [radius=0.16] 
	node[white] {{\fontfamily{qag}\selectfont \tiny ID}};
	\draw[white, fill=white] (-0.0625,0.095) 
	circle [radius=0.007];
	\end{tikzpicture}
	\hspace{-2.5mm}
 }
 \foreach \x in {A, ..., Z}{%
 	\expandafter\xdef\csname orcid\x\endcsname{\noexpand\href{https://orcid.org/\csname orcidauthor\x\endcsname}{\noexpand\orcidicon}}
 }

  \foreach \x in {a, ..., z}{%
  	\expandafter\xdef\csname orcid\x\endcsname{\noexpand\href{https://orcid.org/\csname orcidauthor\x\endcsname}{\noexpand\orcidicon}}
 }

\begin{document}

   \title{TOI-4336~A~b: A temperate sub-Neptune ripe for atmospheric characterization in a nearby triple M-dwarf system}


   \author{M.~Timmermans\orcidP{}\inst{\ref{astro_liege}}
          \and
          G.~Dransfield\orcidA{}\inst{\ref{Birm_Univ}}
          \and 
          M.~Gillon\orcidC{}\inst{\ref{astro_liege}}
          \and
          A.H.M.J.~Triaud\orcidB{}\inst{\ref{Birm_Univ}}
          \and B.~V.~Rackham\orcidK{}\inst{\ref{Depart_Earth_MIT},\ref{Kavli_MIT}} 
          \and
          C.~Aganze\inst{\ref{UCSDiegoP}, \ref{StanfordKIPAC}} 
          \and K.~Barkaoui\orcidD{}\inst{\ref{astro_liege},\ref{Depart_Earth_MIT},\ref{IAC}} 
          \and
          C.~Brice\~{n}o\orcidb{}\inst{\ref{CTIO_Obs}} 
          \and
          A.~J.~Burgasser\orcidJ{}\inst{\ref{UCSDiegoA}} 
          \and 
          K.A.~Collins\orcidM{}\inst{\ref{Harvard_USA}} 
          \and
          M.~Cointepas\inst{\ref{Univ_Grenoble},\ref{Dep_Astro_Geneve}} 
          \and
          M.~Dévora-Pajares\inst{\ref{Granada}} 
          \and
          E.~Ducrot\orcidI{}\inst{\ref{LESIA},\ref{AIM_CEA_CNRS}} 
          \and
          S.~Z\'u\~niga-Fern\'andez\orcidF{}\inst{\ref{astro_liege}} 
          \and
          S.B.~Howell\orcidG{}\inst{\ref{Aus_St_Univ}} 
          \and
           L.~Kaltenegger\orcidV{}\inst{\ref{Carl_Sag_Univ}} 
          \and
          C.A.~Murray\orcidY{}\inst{\ref{Colorado}} 
          \and
           E.K.~Pass\orcidZ{}\inst{\ref{Harvard_USA}} 
          \and
          S.N.~Quinn\orcidq{}\inst{\ref{Harvard_USA}} 
          \and 
          S.N.~Raymond\orcida{}
          \inst{\ref{Lab_Astro_Bordeaux}} 
          \and
          D.~Sebastian\orcidH{}\inst{\ref{Birm_Univ}} 
          \and 
          K.G.~Stassun\orcidE{}\inst{\ref{Vand_Univ}} 
          \and
          C.~Ziegler\inst{\ref{Aus_St_Univ}} 
          \and 
          J.~M.~Almenara\orcidW{}\inst{\ref{Univ_Grenoble}} 
          \and 
          Z.~Benkhaldoun\orcidl{}\inst{\ref{ouka}} 
          \and 
          X.~Bonfils\inst{\ref{Univ_Grenoble}} 
          \and
          J.~L.~Christiansen\orcidj{}\inst{\ref{Caltech}} 
          \and
          F.~Davoudi\orcidQ{}\inst{\ref{astro_liege}} 
          \and
          J.~de Wit\orcidz{}\inst{\ref{Depart_Earth_MIT}} 
          \and 
          L.~Delrez\orcidO{}\inst{\ref{astro_liege},\ref{STAR_Uliege}} 
          \and
          B.-O.~Demory\inst{\ref{unibe}} 
          \and
          W.~Fong\inst{\ref{Kavli_MIT}} 
          \and
          G.~F{\H u}r{\' e}sz\inst{\ref{Kavli_MIT}} 
          \and
          M.~Ghachoui\inst{\ref{astro_liege},\ref{ouka}} 
          \and 
          L.J.~Garcia\orcide{}\inst{\ref{flatiron}} 
          \and
          Y.~G\'omez Maqueo Chew\orcidX{}\inst{\ref{ciudad}} 
          \and
          M.~J.~Hooton\orcidS{}\inst{\ref{Cavendish}} 
          \and 
          K.~Horne\orcidU{}\inst{\ref{St_Andrews}} 
          \and
          M.~N. G{\"u}nther\orcidc{}\inst{\ref{ESA}} 
          \and 
          E.~Jehin\orcidL{}\inst{\ref{STAR_Uliege}} 
          \and
          J.M.~Jenkins\inst{\ref{Ames_NASA}} 
          \and
          N.~Law\inst{\ref{Univ_North_Car}} 
          \and
          A.W.~Mann\inst{\ref{Univ_North_Car}} 
          \and
          F.~Murgas\orcidf{}\inst{\ref{IAC},\ref{la_laguna}} 
          \and
          F.J.~Pozuelos\orcidN{}\inst{\ref{iaa}} 
          \and
          P.P.~Pedersen\orcidd{}\inst{\ref{Zurich}} 
          \and
          D.~Queloz\inst{\ref{Cavendish},\ref{Zurich}} 
          \and 
          G.~Ricker\inst{\ref{Kavli_MIT}} 
          \and
          P.~Rowden\inst{\ref{Royal_asto_society_london}} 
          \and 
          R.~P.~Schwarz\orcidT{}\inst{\ref{Harvard_USA}} 
          \and
          S.~Seager\inst{\ref{Kavli_MIT},\ref{la_laguna},\ref{Univ_of_Maryl}} 
          \and
          R.~L.~Smart\orcidx{}\inst{\ref{oato}} 
          \and
          G.~Srdoc\inst{\ref{Kotizarovci_obs_Croatia}} 
          \and
          S.~Striegel\inst{\ref{SETI_inst}} 
          \and
          S.~Thompson\inst{\ref{Cavendish}}
          \and
          R.~Vanderspek\inst{\ref{Kavli_MIT}} 
          \and
          J.~N.~Winn\inst{\ref{Astro_Prin}}
          }

   \institute{Astrobiology Research Unit, Université de Liège, 19C Allée du 6 Août, 4000 Liège, Belgium \label{astro_liege}
         \and 
         School of Physics \& Astronomy, University of Birmingham, Edgbaston, Birmingham B15 2TT, United Kingdom  \label{Birm_Univ}
         \and
         Department of Earth, Atmospheric and Planetary Science, Massachusetts Institute of Technology, 77 Massachusetts Avenue, Cambridge, MA 02139, USA \label{Depart_Earth_MIT}
        \and 
        Kavli Institute for Astrophysics and Space Research, Massachusetts Institute of Technology, Cambridge, MA 02139, USA \label{Kavli_MIT}
        \and
        Department of Physics, UC San Diego, 9500 Gilman Drive, La Jolla, CA 92093, USA \label{UCSDiegoP}
        \and
        Kavli Institute for Particle Astrophysics \& Cosmology, Stanford University, Stanford, CA 94305, USA \label{StanfordKIPAC}
        \and
        Instituto de Astrofísica de Canarias (IAC), 38205 La Laguna, Tenerife, Spain \label{IAC}
        \and
        Cerro Tololo Inter-American Observatory/NSF's NOIRLab, Casilla 603, La Serena 1700000, Chile \label{CTIO_Obs}
        \and 
        Department of Astronomy \& Astrophysics, UC San Diego, 9500 Gilman Drive, La Jolla, CA 92093, USA \label{UCSDiegoA}
        \and
        Center for Astrophysics \textbar{} Harvard \& Smithsonian, 60 Garden Street, Cambridge, MA, 02138, USA \label{Harvard_USA}
        \and
        Univ. Grenoble Alpes, CNRS, IPAG, 38000 Grenoble, France \label{Univ_Grenoble}
        \and
        Observatoire de Genève, Département d’Astronomie, Université de Genève, Chemin Pegasi 51b, 1290 Versoix, Switzerland \label{Dep_Astro_Geneve}
        \and 
        Dpto. Física Teórica y del Cosmos, Universidad de Granada, 18071, Granada, Spain \label{Granada}
        \and 
        LESIA, Observatoire de Paris, Université PSL, CNRS, Sorbonne Université, Université Paris Cité, 5 place Jules Janssen, 92195 Meudon, France
         \label{LESIA}
        \and
        AIM, CEA, CNRS, Universit\'e Paris-Saclay, Universit\'e de Paris, F-91191 Gif-sur-Yvette, France \label{AIM_CEA_CNRS}
        \and
        Department of Physics \& Astronomy, Vanderbilt University, 6301 Stevenson Center Ln., Nashville, TN 37235, USA  \label{Vand_Univ}
        \and 
        Department of Physics, Engineering and Astronomy, Stephen F. Austin State University, 1936 North St, Nacogdoches, TX 75962, USA \label{Aus_St_Univ}
        \and
        Carl Sagan Institute, Cornell University, 302 Space Science Building, 14850 Ithaca, NY, USA \label{Carl_Sag_Univ}
        \and 
        Department of Astrophysical and Planetary Sciences, University of Colorado Boulder, Boulder, CO 80309, USA \label{Colorado}
        \and
        Laboratoire d'Astrophysique de Bordeaux, CNRS and Universit{\'e} de Bordeaux, All{\'e}e Geoffroy St. Hilaire, 33165 Pessac, France \label{Lab_Astro_Bordeaux}
        \and 
        Oukaimeden Observatory, High Energy Physics and Astrophysics Laboratory, Faculty of sciences Semlalia, Cadi Ayyad University, Marrakech, Morocco \label{ouka}
        \and 
        Caltech/IPAC-NASA Exoplanet Science Institute, 770 S. Wilson Avenue, Pasadena, CA 91106, USA \label{Caltech}
        \and
        Space Sciences, Technologies and Astrophysics Research (STAR) Institute, Université de Liège, Allée du 6 Août 19C, 4000 Liège, Belgium \label{STAR_Uliege}
        \and 
        Center for Space and Habitability, University of Bern, Gesellschaftsstrasse 6, 3012, Bern, Switzerland \label{unibe}
        \and
        Center for Computational Astrophysics, Flatiron Institute, New York, NY, USA \label{flatiron}
        \and 
        Universidad Nacional Aut\'onoma de M\'exico, Instituto de Astronom\'ia, AP 70-264, Ciudad de M\'exico,  04510, M\'exico \label{ciudad}
        \and
        Cavendish Laboratory, JJ Thomson Avenue, Cambridge CB3 0HE, UK \label{Cavendish}
        \and 
        SUPA Physics and Astronomy, University of St. Andrews, Fife, KY16 9SS Scotland, UK \label{St_Andrews}
        \and
        European Space Agency (ESA), European Space Research and Technology Centre (ESTEC), Keplerlaan 1, 2201 AZ Noordwijk, The Netherlands \label{ESA}
        \and
        NASA Ames Research Center, Moffett Field, CA 94035, USA \label{Ames_NASA}
        \and 
        Department of Physics and Astronomy, The University of North Carolina at Chapel Hill, Chapel Hill, NC 27599-3255, USA \label{Univ_North_Car}
        \and
        Departamento de Astrof\'{i}sica, Universidad de La Laguna (ULL), 38206 La Laguna, Tenerife, Spain \label{la_laguna}
        \and
        Instituto de Astrof\'isica de Andaluc\'ia (IAA-CSIC), Glorieta de la Astronom\'ia s/n, 18008 Granada, Spain \label{iaa}
        \and
        ETH Zurich, Department of Physics, Wolfgang-Pauli-Strasse 2, CH-8093 Zurich, Switzerland \label{Zurich}
        \and 
        Royal Astronomical Society, Burlington House, Piccadilly, London W1J 0BQ, UK \label{Royal_asto_society_london}
        \and 
        Department of Astronomy, University of Maryland, College Park, MD 20742, USA \label{Univ_of_Maryl}
        \and 
        Istituto Nazionale di Astrofisica, Osservatorio Astrofisico di Torino, Strada Osservatorio 20, I-10025 Pino Torinese, Italy \label{oato}
        \and 
        Kotizarovci Observatory, Sarsoni 90, 51216 Viskovo, Croatia \label{Kotizarovci_obs_Croatia}
        \and 
        SETI Institute, Mountain View, CA 94043 USA/NASA Ames Research Center, Moffett Field, CA 94035 USA \label{SETI_inst}
        \and 
        Department of Astrophysical Sciences, Princeton University, Princeton, NJ 08544, USA \label{Astro_Prin}
            }

   \date{Received {...}; accepted {...}} 

 
  \abstract
   {Small planets transiting bright nearby stars are essential to our understanding of the formation and evolution of exoplanetary systems. However, few constitute prime targets for atmospheric characterization, and even fewer are part of multiple star systems.   
    }
   {This work aims to validate TOI-4336~A~b, a sub-Neptune-sized exoplanet candidate identified by the \tess space-based transit survey around a nearby M-dwarf.}
   {We validate the planetary nature of TOI-4336~A~b through the global analysis of TESS and follow-up multi-band high-precision photometric data from ground-based telescopes, medium- and high-resolution spectroscopy of the host star, high-resolution speckle imaging, and archival images.}
   {The newly discovered exoplanet TOI-4336~A~b has a radius of 2.1$\pm$0.1R$_{\oplus}$. Its host star is an {M3.5}-dwarf star of mass 0.33$\pm$0.01M$_{\odot}$ and radius 0.33$\pm$0.02R$_{\odot}$ member of a hierarchical triple M-dwarf system 22 pc away from the Sun.
   The planet's orbital period of 16.3 days places it at the inner edge of the Habitable Zone of its host star, the brightest of the inner binary pair.  The parameters of the system make TOI-4336~A~b an extremely promising target for the detailed atmospheric characterization of a temperate sub-Neptune by transit transmission spectroscopy with \jwst.}
   {}

   \keywords{planets and satellites: detection -
             (stars:) planetary systems -
             stars: low-mass - 
             techniques: photometric -
             planets and satellites: individual: TOI-4336~A~b
               }

   \maketitle
%

\section{Introduction}



The Transiting Exoplanet Survey Satellite \citep[\tess,][]{tess} has already added about 400 confirmed planets to the known sample\footnote{NASA Exoplanet Archive, Jun 2023, https://exoplanetarchive.ipac.caltech.edu/}. More than half of these new objects are smaller than Neptune, and transit stars bright enough for detailed characterization with high-precision spectroscopy. Additionally, about 70 of these planets are hosted by M dwarfs. While the \kepler mission revealed a large abundance of such planets with no equivalent in the solar system \citep{ 2011_borucki_occurence_rate,dressing2013_occurence_rate}, most of the {\kepler} planets are currently out of reach {for a detailed} characterization due to the faintness of their host star. 



The study of the bulk composition of these small planets shows two distinct populations having radii smaller than that of Neptune and bigger than the Earth's. One, {"super-Earths"}, are thought to be rocky planets with either thin or no atmosphere, while the other, {"mini-Neptunes"}, show smaller densities consistent with an extended atmosphere or a significant water fraction \citep[e.g.][]{2015ApJ_Rogers_subNept,2008ApJ_water_world_or_atm}. The formation pathways of these planets are not fully understood \citep[for a comprehensive review see][]{bean21}, especially the paucity of planets found between $\sim 1.5$ and $\sim 2.5$ R$_{\oplus}$ \citep{Fulton2017} for FGK stars. Several theories such as gas-poor formation \citep{Lee2014}, and atmospheric loss either by photoevaporation \citep{owen2013} or core-powered mass loss \citep{gupta2019} could explain {this} so-called {"radius valley"}, which may drift towards smaller radii for low-mass stars \citep{Berger2020,cloutier2020}, depending on the model. Another approach considers a density valley rather than a radius valley for M-type stars \citep{Luque2022}. The study distinguishes three populations: rocky planets, water worlds, and puffy mini-Neptunes, with a common formation history for these last two. To illuminate these different formation theories, an increase in the sample of {thoroughly} characterized sub-Neptune-sized planets is critical. 
Small and cool M dwarfs are particularly interesting targets for transmission spectroscopy as they present high signal-to-noise ratios (S/N) even for smaller transiting planets. Extended atmospheres, like those expected for mini-Neptunes, should further increase the S/N for transmission spectroscopy. 

{TOI-4336~A is part of a {hierarchical} triple M-dwarf system {(M3.5, M3.5 and M4-type stars)} located at 22 pc from the Sun. The host star is the brightest component of the inner binary pair which has a minimum orbital separation of over a hundred au, { and is at a projected angular separation of 6.25\arcsec.} {We adopt them as TOI-4336~A and B (TIC\,166184428 and TIC\,166184426, respectively) according to their brightness.} The third star, which we will refer to as TOI-4336~C (TIC\,166184390), is the latest of the system and is on a wider orbit at a distance of over 2900 au, {and at an angular separation of 98.44\arcsec}. Here we report the detection of TOI-4336~A~b, a planet {that} lies at the inner edge of the empirical Habitable Zone (HZ) \citep{kopparapu2013, kopparapu2017, kaltenegger2017, 2013_HZ_limits_Julien} of the system. It receives less irradiation than a young Venus, but since M-dwarf irradiation warms planets more effectively than Sun-like stars (see, e.g., \citealt{kasting1993}), that flux moves TOI-4336~A~b just {closer to the star than} the inner edge of the empirical HZ, making the planet an intriguing example of a planet near a boundary of the HZ.} 

We describe in Section \ref{sec:stellar_charact} the observations and methods used to characterise the triple star system. 
The \tess and ground-based observations used to validate the planetary nature of TOI-4336~A~b are detailed in Section \ref{sec:photo_data}, and the statistical validation is reported in Section \ref{sec:validation}. The joint analysis of all the photometric data is presented in Section \ref{sec:global_analysis} and the results are discussed in Section \ref{sec:discussion}. {Finally, we present our conclusions in Section \ref{sec:conclusions}.}



\section{Stellar characterization}
\label{sec:stellar_charact}
\subsection{Spectroscopic reconnaissance}



\begin{figure*}[hbt!]
    \centering
    \includegraphics[width=\textwidth]{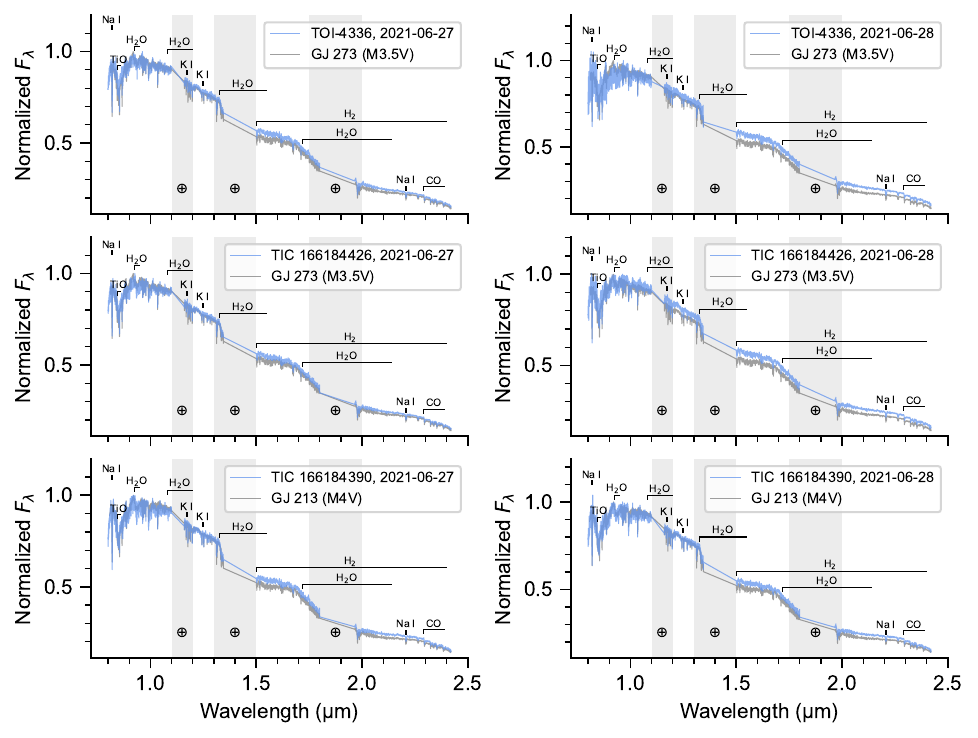}
    \caption{ {
        SpeX spectra of TOI-4336 (TOI-4336~A, top row), TIC\,166184426 (TOI-4336~B, middle row), and TIC\,166184390 (TOI-4336~C, bottom row) from 2021~Jun~27 (left column) and 2021~Jun~28 (right column). The target spectrum (blue) is shown alongside the best-fit spectral template from the IRTF Spectral Library (grey). All spectra are normalized to their flux in the 0.9--1.4\,$\mu$m region. Wavelengths with strong telluric absorption are shaded (and largely masked from the spectra), and prominent M-dwarf absorption features are highlighted.}
    }
    \label{fig:spex}
\end{figure*}

We gathered near-infrared spectra of TOI-4336~A and the two resolved, co-moving stars (TOI-4336~B and TOI-4336~C) with the SpeX spectrograph \citep{Rayner2003} on the 3.2-m NASA Infrared Telescope Facility (IRTF) on 2021\,Jun\,27 (UT) and again on 2021\,Jun\,28 (UT).
{We used the short-wavelength cross-dispersed (SXD) mode of SpeX and the $0\farcs5 \times 15\arcsec$ slit aligned to the parallactic angle, which yielded spectra covering 0.75--2.42\,$\mu$m at $R{\sim}1200$ (Figure \ref{fig:spex}). We collected 6 exposures of each target, nodding in an ABBA pattern. Our total exposure times were 360\,s each for TOI-4336~A and {TOI-4336~B}, and 540\,s for TIC\,166184390 {TOI-4336~C}. We collected the standard set of SXD flat-field and arc-lamp exposures immediately after the science frames, followed by the A0\,V standard HD\,130163.}
We reduced the data with SPEXTOOL v4.1 \citep{Cushing2004}.
We used SPLAT to compare the spectra to standards in the {IRTF Spectral Library \citep{Cushing2005, Rayner2009}, focusing on the 0.9--1.4\,$\mu$m region for the spectral classification \citep{Kirkpatrick2010},} and to estimate {metallicity} [Fe/H] {via the \citet{2013AJ....145...52M} relation} \citep[see][for details]{speculoos2}.
{Between nights, the spectral-type determinations of each target are identical and their metallicity estimates are consistent at ${<}1\sigma$.}
We estimate spectral types of {M3.5$\pm$0.5 for TOI-4336~A and TOI-4336~B and M4.0$\pm$0.5 for TOI-4336~C.}
{Combining measurements} from both nights, {we obtain final [Fe/H] estimates of $-0.20\pm0.12$, $-0.21\pm0.12$, and $-0.17\pm0.12$} for TOI-4336~A, B, and C, respectively {(see Table \ref{tab:properties_TOI-4336}). }

\begin{figure}[hbt]
	\centering
	\includegraphics[width=\columnwidth]{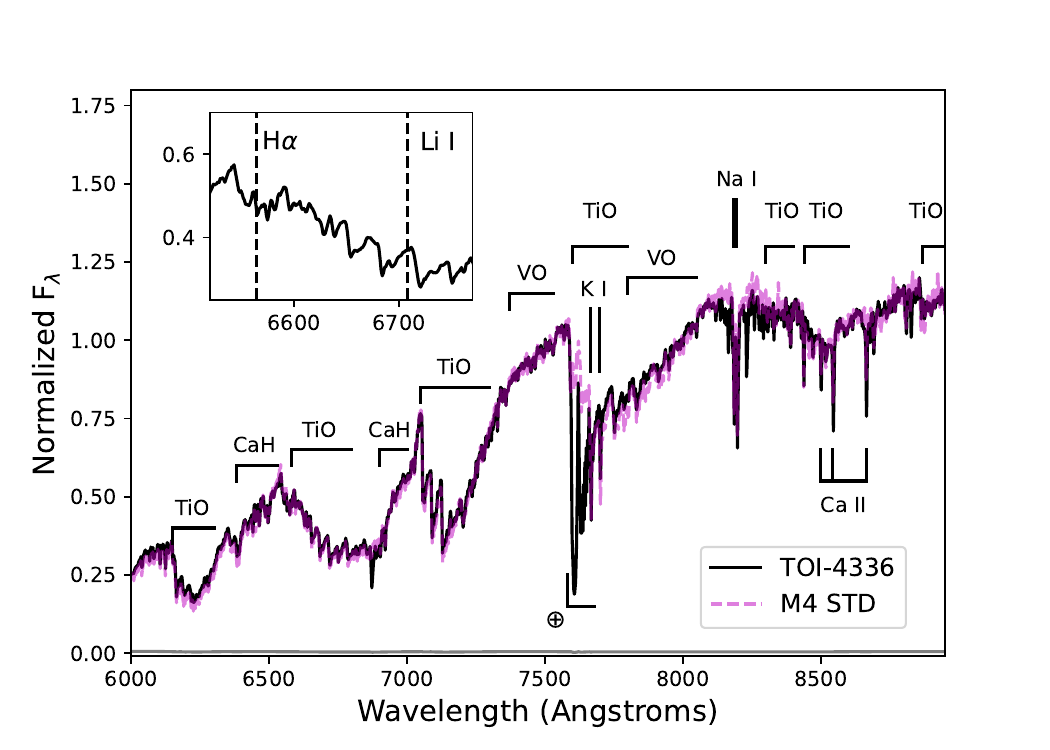}
	\caption{{LDSS3 red optical spectrum of TOI-4336~A (black line), compared to its best-fit M4 template \cite[magenta line]{2017ApJS..230...16K}. Spectra are normalized in the 7400--7500~{\AA} region, and major absorption features are labeled, including regions of strong telluric absorption ($\oplus$). The inset box shows a close-up of the region encompassing H$\alpha$ (6563~{\AA}) and Li\,\textsc{i}~ (6708~{\AA}) features, neither of which is detected.}}
	\label{fig:LDSS3_spectrum}
\end{figure}

We acquired an optical spectrum of TOI-4336~A on 2022\,Jan\,07 (UT) using the Low Dispersion Survey Spectrograph \citep[LDSS3-C,][]{Stevenson2016} on the 6.5-m \textit{Magellan} II (Clay) Telescope under clear and stable conditions.
{We used LDSS-3C in long-slit mode with the standard setup (fast readout speed, low gain, $1{\times}1$ binning) and the VPH-Red grism, OG-590 blocking filter, and the $0\farcs75 \times 4'$ center slit, which provides spectra covering 6000--10\,000\,\AA{} at $R{\sim} 1810$.
We collected eight, 60-s exposures of the target, followed by a 1-s arc-lamp exposure and three 10-s flat field exposures with the quartz high lamp and reduced the data with a custom Python-based pipeline \citep{Dransfield2023}.
We used the ratio of the spectrum of the G2\,V star HR\,5325, observed at a similar airmass, to a G2\,V template from \cite{1998PASP..110..863P} to compute a relative flux calibration of the TOI-4336 spectrum. No correction was made to address telluric absorption.}
The reduced spectrum is shown in {Figure \ref{fig:LDSS3_spectrum}}. It was then analyzed using tools in the \texttt{kastredux} package\footnote{\url{https://github.com/aburgasser/kastredux}.}.
Comparison with the Sloan Digital Sky Survey templates from \cite{2017ApJS..230...16K} shows an excellent match to an M4 dwarf template. This is confirmed by analysis of spectral classification indices from \cite{1995AJ....110.1838R}; \cite{Lepine2003}; and \cite{2007MNRAS.381.1067R}. We also computed the $\zeta$ metallicity index \citep{2007ApJ...669.1235L,2013AJ....145..102L}, determining a value of 1.026$\pm$0.002, corresponding to a metallicity of [Fe/H] = $+$0.04$\pm$0.20 based on the empirical calibration of \cite{2013AJ....145...52M}. 

We obtained high-resolution spectroscopic observations with the CHIRON spectrograph, {located on the SMARTS 1.5-meter telescope at Cerro Tololo Inter-American Observatory}, \citep{2013PASP..125.1336T} for both TOI-4336~A and the co-moving companion TOI-4336~B. 
We used CHIRON's `slicer' mode, which employs an image slicer to achieve a resolving power of $R\sim80,000$ from $4100$\ to $8700\,\AA{}$. {Our observations yield per-pixel signal-to-noise ratios that range from 10--12 in the TiO bands around
7100 \AA.} Spectra were extracted using the official CHIRON pipeline \citep{2021AJ....162..176P}, and we derived radial velocities (RVs) and spectral line profiles as described by \citet{2023AJ....166...11P}\footnote{This analysis builds upon the \texttt{tres-tools} package: \url{https://github.com/mdwarfgeek/tres-tools}.}. {This reduction is specifically optimized for mid-to-late M dwarfs and produces carefully calibrated relative RVs. For our relative RVs, the error budget is dominated by spectrograph stability. For our absolute RVs, the dominant uncertainty is a 0.5kms$^{-1}$ error in the RV scale; this error stems from the absolute RV uncertainty in a comparison spectrum of Barnard's Star, derived using 17 years of measurements from the CfA Digital Speedometer. The RV measurements are given in Table \ref{tab:properties_TOI-4336}.} TOI-4336~A was observed on 2021\,Feb\,06 and 2021\,Jul\,12, and TOI-4336~B was observed on 2021\,Feb\,06.

We find no evidence for significant velocity variation in TOI-4336~A, and the difference in absolute radial velocity between the two stars is consistent with orbital motion in a wide binary given their separation. We detect no significant rotational broadening for either star, setting upper limits of $v\sin{i_\star} < 1.9\ {\rm km\,s}^{-1}$. {This limit corresponds to half a resolution element of the CHIRON spectrograph.} H$\alpha$ {is seen in absorption for both stars}, and following \citet{2017ApJ...834...85N}, we measure ${\rm EW}_{{\rm H}\alpha} = 0.1644 \pm 0.0025${\AA} for TOI-4336~A, which places it among the sequence of quiescent M dwarfs. Taken together, we interpret the lack of detectable activity and rotational broadening as an indication that the star is not young, which is in agreement with the results of the LDSS3 spectral analysis, {this is discussed in Section \ref{sec:Age_estimate}.}

\subsection{Spectral Energy Distribution}
\label{sec:sed}

\begin{figure}[hbt]
\centering
\begin{minipage}{0.5\textwidth}
\includegraphics[width=\textwidth]{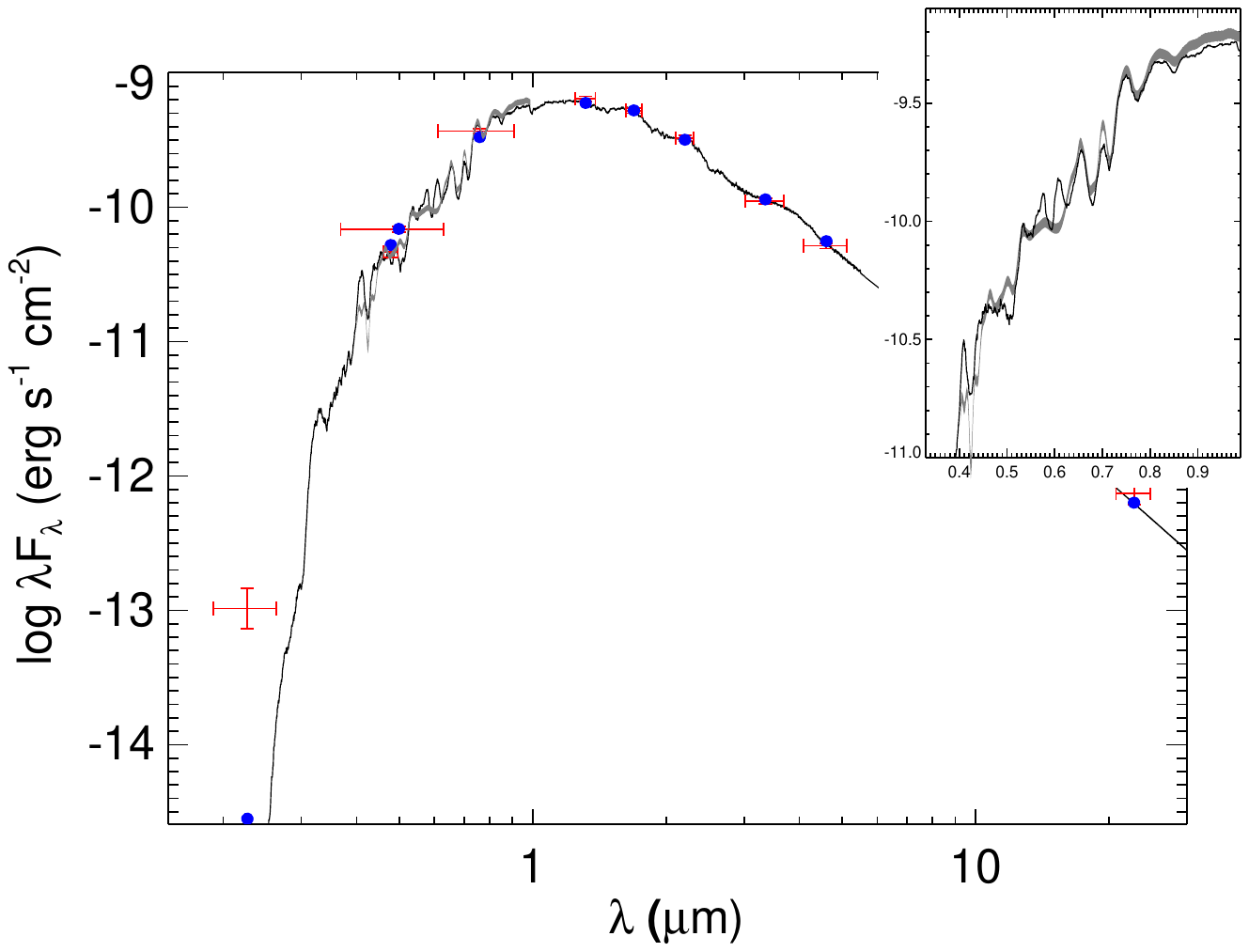}
\end{minipage}

\vspace{-4em}
\centering
\begin{minipage}{0.5\textwidth}
\includegraphics[width=\textwidth]{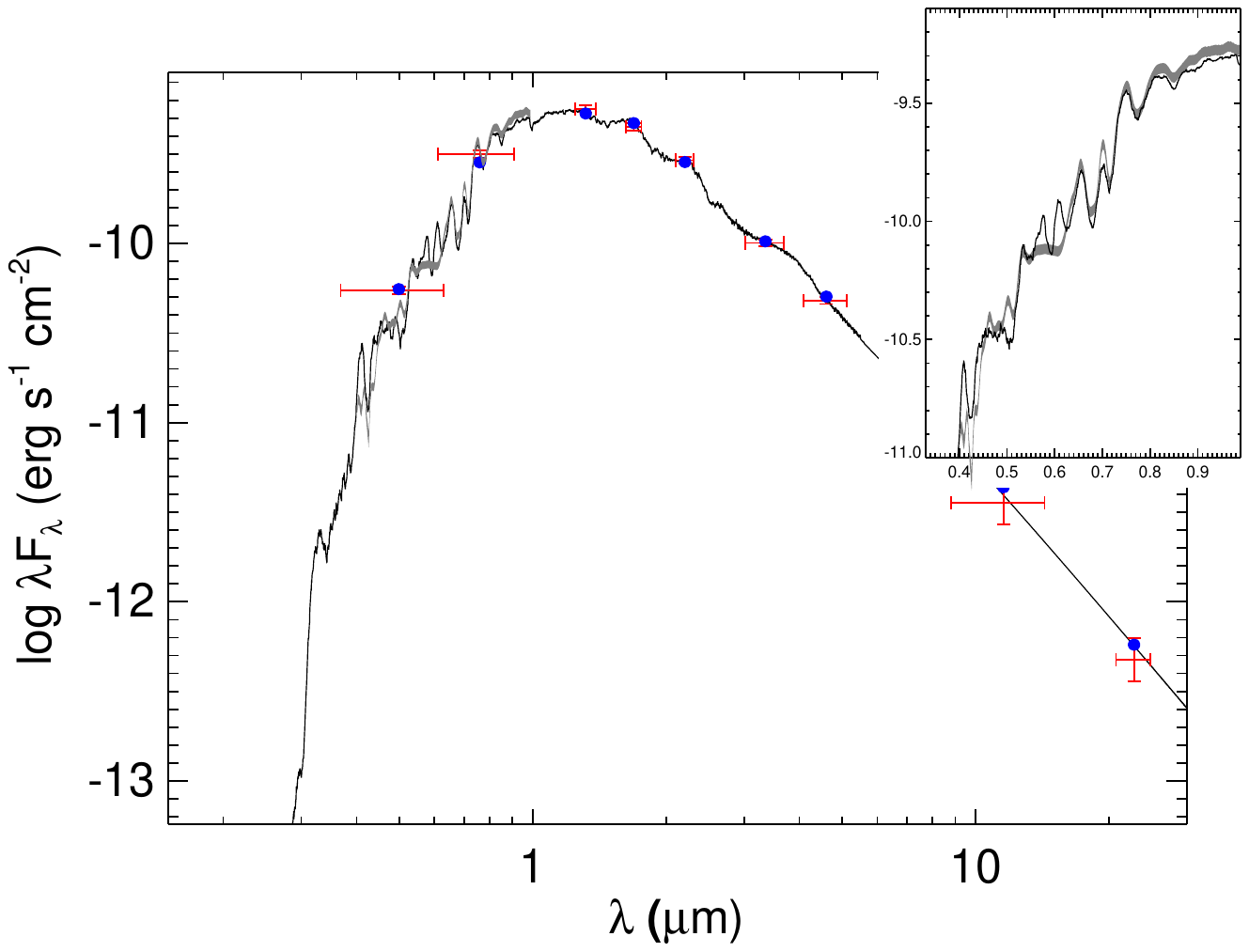}
\end{minipage}

\vspace{-4em}
\centering
\begin{minipage}{0.5\textwidth}
\includegraphics[width=\textwidth]{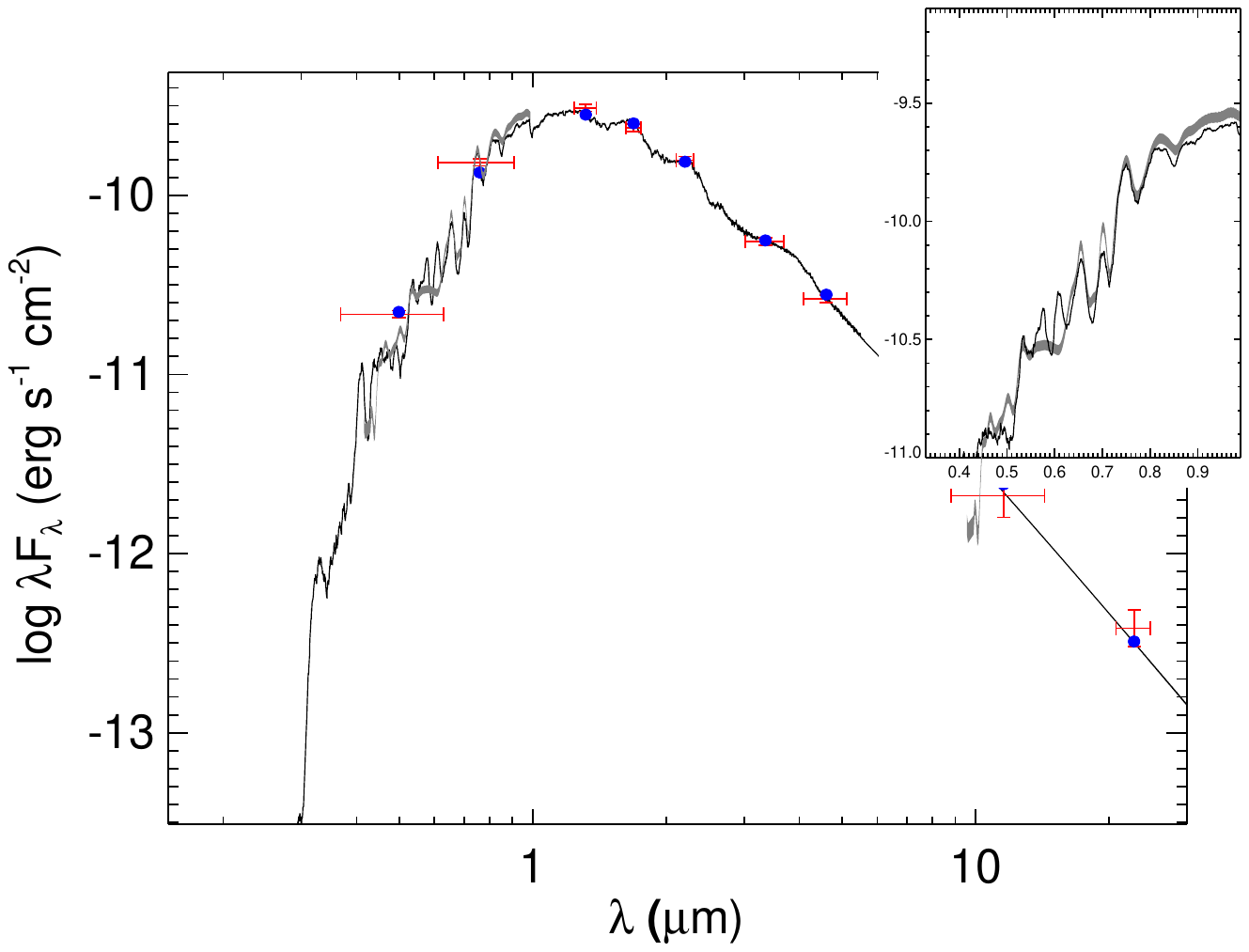}
\end{minipage}
\caption{{Spectral energy distribution of {TOI-4336~A} {(top panel)} and its companion stars {TOI-4336~B} {(middle panel)} and {TOI-4336~C} {(bottom panel)}. Red symbols represent the observed photometric measurements, the horizontal bars represent the effective width of the passband. Blue symbols are the model fluxes from the best-fit PHOENIX atmosphere model (black). The {\it Gaia\/} spectrophotometry is represented as a grey swathe; a closeup view is shown in the inset. 
}}
\label{SED_plots}
\end{figure}

As an independent determination of the basic stellar parameters, we performed an analysis of the broadband spectral energy distribution (SED) of the star together with the {\it Gaia\/} DR3 parallax (with no systematic offset applied; see, e.g., \citealt{StassunTorres:2021}), 
following the procedures described in \cite{Stassun:2016,Stassun:2017,Stassun:2018}. {We pulled the $JHK_S$ magnitudes from {\it 2MASS} {\citep{2mass_catalog}}, the W1--W4 magnitudes from {\it WISE} {\citep{wisecat}}, the $G_{\rm BP}$ {and} $G_{\rm RP}$ magnitudes from {\it Gaia} {\citep{gaiaDR3cat}}, and the NUV flux from {\it GALEX} {\citep{2017_galex_bianchi}}. We also used the {\it Gaia\/} spectrophotometry spanning 0.4--1.0~$\mu$m, providing an especially strong constraint on the overall absolute flux calibration. }
Together, the available photometry spans the stellar SED over the wavelength range 0.2--22~$\mu$m (see Figure~\ref{SED_plots}).  

We performed a fit using PHOENIX stellar atmosphere models \citep{Husser:2013}, with the effective temperature ($T_{\rm eff}$) and metallicity ([Fe/H]) as free parameters (the surface gravity, $\log g$, has very little influence on the broadband SED). We set the extinction, $A_V$, to zero given the close proximity of the system. The resulting fit (Figure~\ref{SED_plots}) has a reduced $\chi^2$ of 1.3 (excluding the NUV measurement, which suggests some chromospheric activity), with best-fit $T_{\rm eff} = 3300 \pm 75$~K and [Fe/H] = $0.0 \pm 0.2$. The derived parameters are shown in Table \ref{tab:properties_TOI-4336}.

{Integrating the model SED gives the bolometric flux at Earth, $F_{\rm bol} = 7.40 \pm 0.17 \times 10^{-10}$ erg~s$^{-1}$~cm$^{-2}$. Taking the $F_{\rm bol}$ together with the {\it Gaia\/} parallax directly gives the luminosity, $L_{\rm bol} = 0.01163 \pm 0.00027$~L$_\odot$. Similarly, the $F_{\rm bol}$ together with the $T_{\rm eff}$ and the parallax gives the stellar radius, $R_\star = 0.330 \pm 0.015$~R$_\odot$. The stellar mass can also be estimated via the empirical $M_K$ based relations of \cite{Mann:2019}, giving $M_\star = 0.331 \pm 0.010$~M$_\odot$. 
All uncertainties are propagated in the usual manner, except for the $M_K$-based mass estimate {for which} we adopt the systematic uncertainties quoted for the \citet{Mann:2019} relations as the dominant source of error.
These parameters are summarized in Table~\ref{tab:param}.}


For completeness, we applied the same SED-fitting procedures to the other two stars in the system, with the results shown in Figure~\ref{SED_plots} and summarized in Table~\ref{tab:properties_TOI-4336}. {We placed the three stars of the TOI-4336 system in a color-magnitude diagram to compare their properties to nearby M dwarfs (see Figure \ref{fig:color-mag}). TOI-4336~C appears less luminous than the other two, which is consistent with their spectral types.} 

\begin{table*}[t] 
\centering
\caption{Properties of the TOI-4336 system.}
\label{tab:properties_TOI-4336}
\begin{tabular}{@{}lcccc@{}}
\toprule
\toprule
{Parameters} & \multicolumn{3}{c}{{Values}} & {Refs}\\
\toprule
\toprule
\textit{Designation} \\
This work  & TOI-4336~A & TOI-4336~B & TOI-4336~C \\
\vspace{0.12cm}
TIC & 166184428 & 166184426 & 166184390 & {[1]}\\
2MASS & J13442546-4020155 & J13442500-4020122 & J13442755-4018400  & {[2]}\\
{\it Gaia\/} (DR3) & 6113245033656232448 & 6113245033659187200 & 6113271494953274752 & {[3]} \\
UCAC4 & 249-060094 & 249-060092 & 249-060096 & {[4]} \\
WISE & J134425.61-402014.8 & J134425.16-402011.4 & J134427.70-401839.2 & {[5]} \\
WDS & J13444-4020A & J13444-4020C & J13444-4020B & {[6]}\\
\textit{Photometric magnitudes} \\
TESS (mag) & 
$11.0196 \pm 0.0074$& 
$11.1825 \pm 0.0074$& 
$11.9501 \pm 0.0073$&
{[1]} \\
B (mag)  & 
$14.489 \pm 0.001$ & 
$14.510 \pm 0.037$ & 
$16.369 \pm  0.012$&
{[1,4,7]}\\
V (mag) & 
$12.893 \pm 0.006$& 
$12.911 \pm 0.052$ & 
$14.732 \pm 0.001$&
{[1,4,7]}\\
G (mag) & 
$12.245831 \pm 0.002774$& 
$12.434232 \pm 0.002785$& 
$13.2893 \pm 0.000731$&
{[3]}\\
J (mag)  & 
$9.453 \pm 0.024$& 
$9.589 \pm 0.024$& 
$10.249 \pm 0.022$&
{[2]}\\
H (mag)  & 
$8.867 \pm 0.046$& 
$9.038 \pm 0.025$& 
$9.723 \pm 0.023$&
{[2]}\\
K (mag)  & 
$8.632 \pm 0.024$& 
$8.756 \pm 0.021$& 
$9.426 \pm 0.021$&
{[2]}\\
W1 (mag) & 
$8.479 \pm 0.029$& 
$8.588 \pm 0.03$& 
$9.243 \pm 0.023$&
{[8]}\\
W2 (mag) & 
$8.333 \pm 0.025$& 
$8.411 \pm 0.026$& 
$9.063 \pm 0.02$ &
{[8]}\\
W3 (mag) & 
$8.218 \pm 0.025$& 
$8.311 \pm 0.027$& 
$8.887 \pm 0.025$&
{[8]}\\
W4 (mag) & 
$7.865 \pm 0.2$& 
$8.345 \pm 0.303$& 
$8.579001 \pm 0.255 $&
{[8]}\\
\textit{Astrometric properties} \\
Distance (pc)  & 
$22.45 \pm 0.02$& 
$22.44 \pm 0.03$& 
$22.45 \pm 0.08$&
{[9]}\\
Parallax (mas) & 
$44.53\pm0.03$ & 
$44.55\pm0.04$ &
$44.50\pm0.02$ &
{[3]} \\
RA (J2000)  &
$13:44:25.4773$ &
$13:44:25.0160$& 
$13:44:27.5674$&
{[3]}\\
DEC (J2000)  &
 $-40:20:15.5222$ &
$-40:20:12.1623$ & 
$-40:18:40.0242$&
{[3]}\\
$\mu_{\rm{RA}}$ (mas yr$^{-1}$) & 
$151.813 \pm 0.033$ &
$150.407 \pm 0.039$& 
$151.991 \pm 0.016$&
{[3]}\\ 
$\mu_{\rm{DEC}}$ (mas yr$^{-1}$)  & 
$68.402 \pm 0.025$ &
$71.661 \pm 0.028$& 
$71.798 \pm 0.014$ &
{[3]}\\ 
RUWE & 
1.86 & 
1.80 & 
1.21 & 
{[3]} \\
{ U (km s$^{-1}$)} &  { 23.5 $\pm$ 0.3} 
&  {23.3$\pm$ 0.3}
&   {24.0$\pm$ 0.9}  & {[3]}\\
{ V (km s$^{-1}$)} &  { 0.6 $\pm$ 0.3} 
&  {0.6 $\pm$ 0.4}
&   {0.2$\pm$ 0.9}  & {[3]}\\
{ W (km s$^{-1}$)} &  { 10.0 $\pm$ 0.2} 
&  {10.4 $\pm$ 0.2}
&   {10.6$\pm$ 0.5}  & {[3]}\\
{Radial velocity (km s$^{-1}$)} & 
{$18.37 \pm 0.36$} &
{$18.36 \pm 0.51$} &
{$19.23 \pm 1.33$} &
{[3]}\\
\textit{{This work}} \\
SpT & 
{M3.5 $\pm$ 0.5} & 
{M3.5 $\pm$ 0.5} & 
{M}4.0 $\pm$ 0.5 & 
SpeX \\ 
$T_{\mathrm{eff}}$ (K) & 
$3300 \pm 75$ & 
$3255 \pm 75$ & 
$3150\pm 75$ & SED 
\\
\vspace{0.12cm}
$[\mathrm{Fe/H}]$ (dex) & 
{$-0.20 \pm 0.12$} & 
{$-0.21 \pm 0.12$} & 
{$-0.17 \pm 0.12$} & 
SpeX 
\\
 & 
{$+0.04 \pm 0.20$} & 
- & 
- & 
LDSS3 
\\
\vspace{0.12cm}
$L_{\rm bol,\star}$ ($L_{\odot}$) & $0.0116 \pm 0.0003$
&  $0.0102 \pm 0.0002$
&  $0.0053 \pm 0.0002$ & $F_{\rm bol,\star}$ + parallax  
\\
\vspace{0.12cm}
$F_{\rm bol}$ ($10^{-10}$ erg\,cm$^{-2}$\,s$^{-1}$)& $7.40 \pm 0.17$
& $6.50 \pm 0.15$
& $3.35 \pm 0.12$
& SED \\
\vspace{0.12cm}
$R_{\star}$ ($R_{\odot}$) &  $0.330 \pm 0.015$
&  $0.318 \pm 0.015$
&   $0.224 \pm 0.012$ & $F_{\rm bol}$+ T$_{\rm{eff}}$ + parallax 
\\
\vspace{0.12cm}
$M_{\star}$ ($M_{\odot}$) &  $0.331 \pm 0.010$
& $0.314 \pm 0.015$
& $0.236 \pm 0.007$ & $M_K$ $^{a}$ 
\\
\vspace{0.12cm}
$\rho_{\star}$ ($\rho_{\odot}$) &  $12.97_{-1.66}^{+2.03}$
& $13.72_{-1.79}^{+2.20}$
& $22.89_{-3.16}^{+3.91}$ 
& $R_{\star} + M_{\star}$
\\
\vspace{0.12cm}
log $g_{\star}$ (cgs) & $4.92 \pm 0.04$
& $4.93 \pm 0.04$
& $5.04_{-0.04}^{+0.05}$ & $R_{\star} + M_{\star}$
\\
\vspace{0.12cm}
Radial velocity (km s$^{-1}$)  & 
$18.0 \pm 0.5$ &  
$18.7 \pm 0.5$&  
- & 
CHIRON \\
\vspace{0.12cm}
{Age (Gyr)} & {6.7$_{-3.1}^{+2.7}$}
& {6.7$_{-3.1}^{+2.7}$}
& {6.7$_{-3.1}^{+2.7}$} & Kinematics \\
\vspace{0.12cm}
\vspace{0.12cm}
\end{tabular}
{References: [1] \cite{2019_stassun_TIC_cat}, [2] \cite{2mass_catalog}, [3] \cite{gaiaDR3cat}, [4]  \cite{ucac4}, [5] \cite{wisecat}, [6] \cite{WDS2001}, [7] \cite{apass_catalog}, [8] \cite{allwise_catalog}, [9] \cite{gaia_distances}}\\
{Notes:}$^{a}$ The stellar mass is estimated via the empirical $M_K$ based relations of \cite{Mann:2019}.
\end{table*}

\subsection{Age estimation}
\label{sec:Age_estimate}

To estimate the age of the system, we compared its kinematics and metallicity to a large sample of stars from GALAH survey DR3 \citep{2018MNRAS.478.4513B}. Ages for stars in the GALAH survey are estimated based on the Bayesian method by \cite{2018MNRAS.473.2004S} {, which is designed to fit the measured distribution of stellar parameters (sky positions, T$_{\rm eff}$, 
$\log$ g, [Fe/H], magnitudes ) with a combination of model isochrones and a Galactic population synthesis model}. {We used the same reference frame as the GALAH DR3 catalog and the Gaia radial velocities to compute the UVW velocities of the system, given in Table \ref{tab:properties_TOI-4336}. }We selected objects with distances $<$~300~pc, and total 3D-velocity $|V-V_{\rm system}|< 10 \,\rm km \,s^{-1}$ and metallicities within 1-standard deviation of the system{, our sample is shown in Figure \ref{fig:age_estimation}}. These constraints yield an age of { 6.7$_{-3.1}^{+2.7}$~Gyr.} Additionally, the lack of {H$\alpha$ (6563~{\AA}) and Li\,\textsc{i}~ (6708~{\AA})} {emission} in any of the optical spectra further confirms that the system is likely old \citep{2017ApJ...834...85N,2021AJ....161..277K}.
{In particular, examining the LDSS3 spectra, we find no evidence of significant H$\alpha$ emission with a 3$\sigma$ equivalent width limit of 0.12~{\AA}, suggesting an age $\gtrsim$4.5~Gyr according to the age--activity relation of \citet{2008AJ....135..785W}. 
However, TOI-4336\,A has a mass {placing it} near the fully convective boundary, and more recent studies have found the average age of transition to the inactive mode is $2.4 \pm 0.3$\,Gyr for fully convective M dwarfs \citep{Medina2022} with some of them transitioning as early as 600\,Myr \citep{Pass2022}.}

{Finally, using the latest version of the BANYAN tool \citep{2018_Gagné_BAYAN}\footnote{\url{https://www.exoplanetes.umontreal.ca/banyan/banyansigma.php}}, we find that the system has a 99.9 percent probability of being a field object, with no association with nearby young moving groups. This further confirms that the triple system is likely old. }

\subsection{Triple M-dwarf system}
\label{sec:orbits}

TOI-4336~A is the primary star of a resolved triple system, with angular separations of 6.25\arcsec and 98.44\arcsec from the B and C components respectively, identified as WDS\,J13444-4020 in the Washington Double Stars Catalog \citep[WDS,][]{WDS2001}. We used the astrometric measurements from WDS, the stellar masses from SED analysis and the {\it Gaia\/} astrometry, distances and proper motions to get a first-order estimation of the orbital elements of the system with \texttt{LOFTI-Gaia} package\footnote{\url{https://github.com/logan-pearce/lofti_gaia}.} \citep{LOFTI_gaia}. We obtained a posterior sample of 10000 accepted orbits for each system (see Appendix \ref{appendix:orbital_fit}). The posterior parameters are given in Table \ref{table:orbital_solution}. The results show that the close binary has a median semi-major axis of 133\,au with a {68\%} confidence interval from 72 to {175\,au. This results correspond to a median periastron separation of 34.7\,au. with a 68\%  confidence interval from 5 to 118\,au. }Although the short phase coverage does not allow a detailed orbital characterization of the system, {these results are used as a point of reference to evaluate the effect of the triple system configuration on the formation of TOI-4336~A~b (see Section \ref{sec:formation in triple system}).}


\section{Photometric observations}
\label{sec:photo_data}
\subsection{TESS}
\label{sec:photo_data_TESS}

The two close components of the system, TOI-4336~A and B, 
have an angular separation of 6\arcsec and thus are located in the same \tess pixel, given the pixel size of 21\arcsec. The tertiary component is, however, resolved in \tess. They were observed in Sector\,11 (2019\,Apr\,26 to 2019\,May\,20), Sector\,38 (2021\,Apr\,29 to 2021\,May\,26) and Sector\,64 (2023\,Apr\,06 to 2023\,May\,04), with {both short- and long- cadence modes. The 2-min cadence image data} were reduced and analyzed using the Science Processing Operations Center \citep[SPOC,][]{SPOC} pipeline at NASA Ames Research Center. 30-min cadence data were obtained by the TESS-SPOC pipeline \citep{TESS-SPOC} and 10-min cadence data were obtained by the Quick Look Pipeline \citep[QLP,][]{QLP}. 

\begin{figure*}[h]
	\centering
	\includegraphics[scale=0.22]{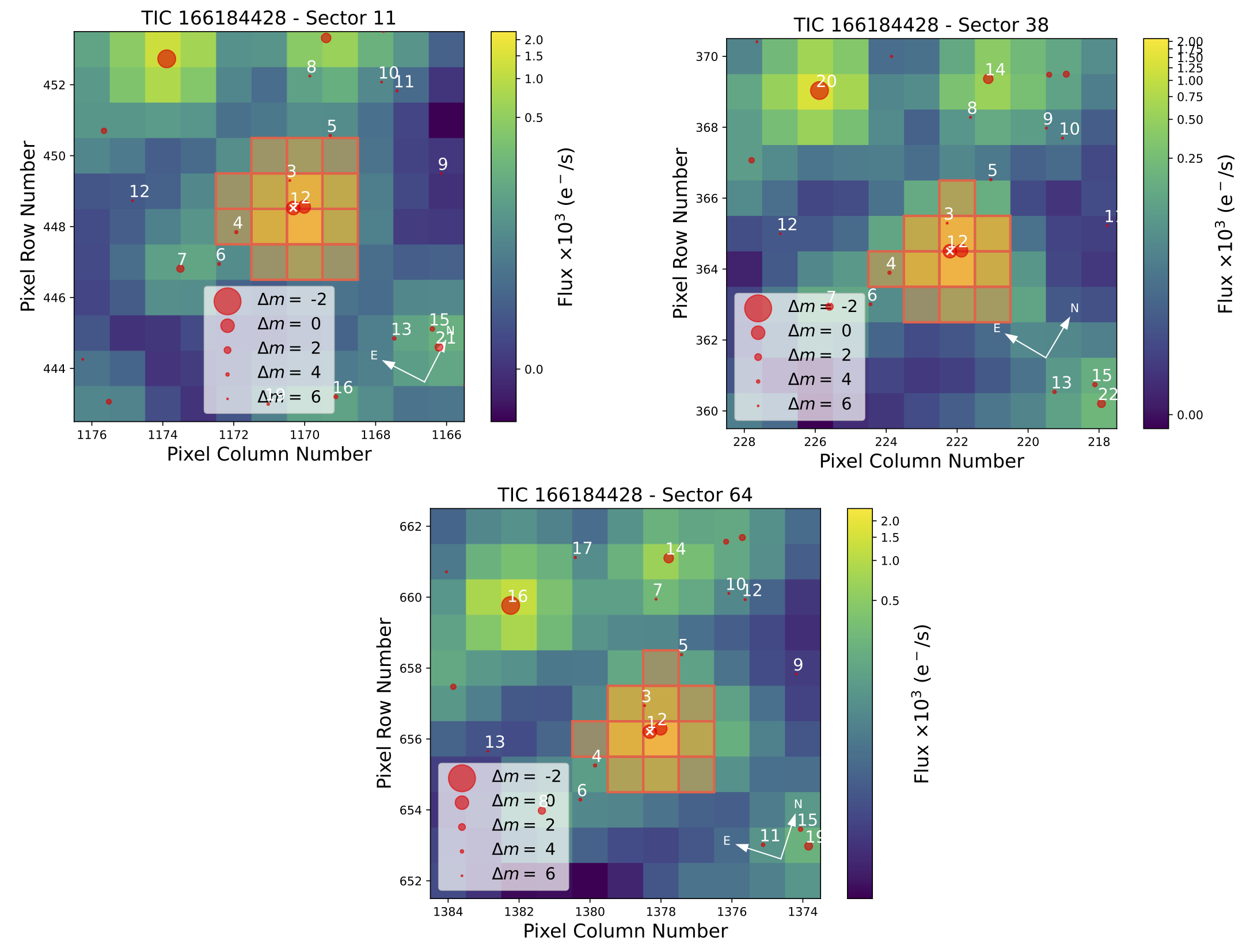}
	\caption{\tess Target Pixel Files showing the custom apertures used to extract the photometry of TOI-4336~A~b in this work for Sectors\,11, 38, and 64.} 
	\label{fig:tpfs}
\end{figure*}

\begin{figure*}
    \centering
    \includegraphics[scale=0.5]{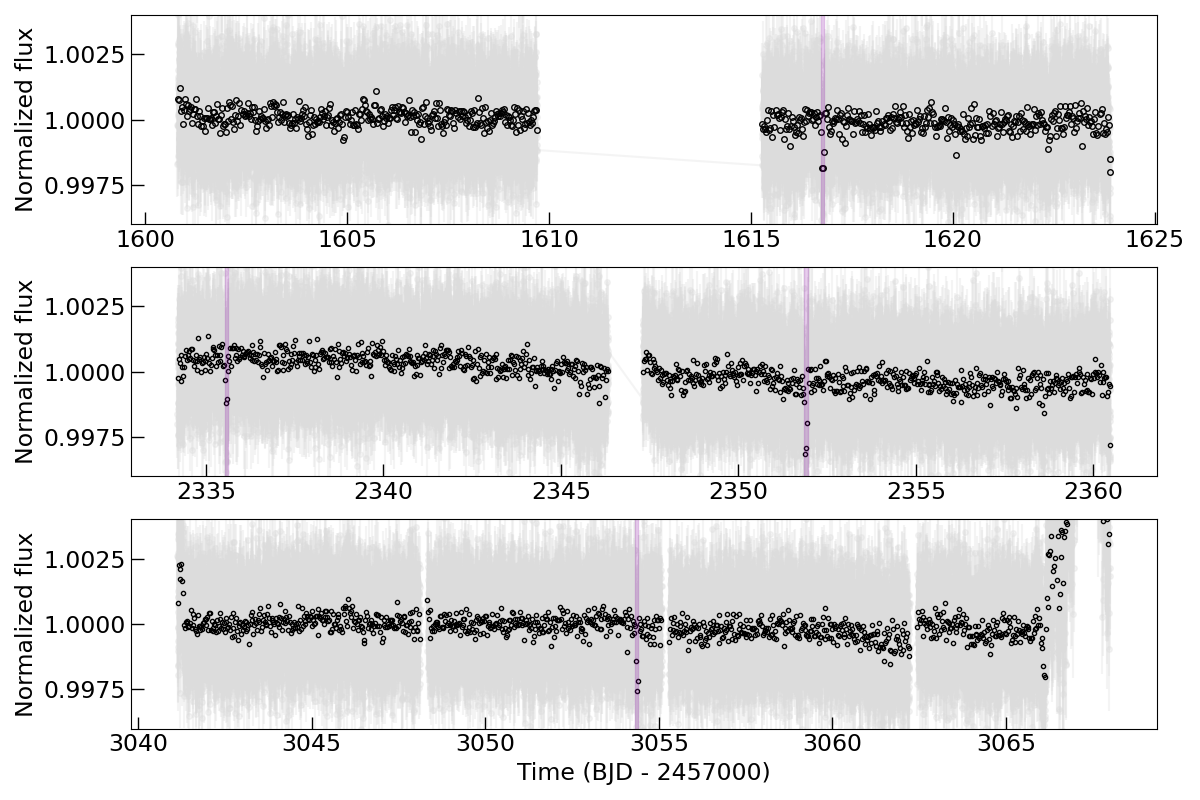}
    \caption{\tess 2-min cadence photometry obtained using custom apertures for Sectors 11, 38, and 64. The transits are highlighted in purple. }
    \label{TESS_sectors}
\end{figure*}

{We extracted the photometry for the global analysis from the TESS Target Pixel Files (TPFs) for all three sectors. 
We retrieved the 2-min \tess TPFs from the Mikulski Archive for Space Telescopes using the \texttt{lightkurve}\footnote{\url{https://github.com/lightkurve/lightkurve}} package \citep{lightkurve} using the hard quality bitmask to remove the images affected by scattered light or sub-optimal attitude control. At the beginning of each orbit of Sector\,11, Camera 1 was subject to scattered light, and the attitude control was disabled for a short period. Using the hard-quality bitmask allows removing this affected bit of data (quality flag 7407). We extracted the photometry from the images using a set of custom apertures inscribed within a square of 4x4 pixels centered on the target, these are shown in Figure \ref{fig:tpfs}. {The \tess TPFs are obtained with \tpfplotter\footnote{\url{https://github.com/jlillo/tpfplotter}} \citep{tpfplotter}.} After the removal of the 5$\sigma$ outliers, we used the Cotrending Basis Vectors (CBVs) obtained by the Presearch and Data Conditioning (PDC) pipeline module of SPOC, a method initially developed to remove low and high-frequency systematic trends in \kepler data \citep{2012_CBVs}. We used the Multi-Scale correction method, which is preferred to preserve transit signals, combined with a correction of short spikes. The correction was calculated using a linear regression approach with an L2 regularization implemented in \texttt{lightkurve}. To prevent under- or over-fitting of the data, the value of the L2 penalty is optimized according to the PDC goodness metrics.  We compared the corrected light curves using the Combined Differential Photometric Precision (CDPP) metric as implemented in \texttt{lightkurve} and selected the one for which the metric was minimal. The extracted light curves are shown in Figure \ref{TESS_sectors}.}

Data Validation Reports were produced by the \tess pipelines each time a new sector became available. The validation tests systematically performed for a planet candidate were passed, although in the first instance the period found was half the true period and the centroid shift analysis did match to TOI-4336~A, TOI-1955.01 was in fact issued for TOI-4336~B. Our ground-based photometric follow-up confirmed the source to be TOI-4336~A. On the \tess side, the SPOC conducted a transit search of Sectors 38 and 64 on 2021\,Jul\,2 and 2023\,Jun\,20 respectively. The transit search was conducted with an adaptive, noise-compensating matched filter \citep{2002_Jenkins,2010_Jenkins,2020_Jenkins}, producing a {threshold crossing event} for which an initial limb-darkened transit model was fitted \citep{2019_Li} and a suite of diagnostic tests were conducted to help make or break the planetary nature of the signal \citep{2018_Twicken}. The TESS Science Office (TSO) reviewed the vetting information and issued an alert for TOI-4336.01 on 2021\,Jul\,28 \citep{2021_Guerrero}. According to the difference image centroiding tests, the host star is located within 4.0 +- 2.7 \arcsec of the transit signal source of TOI-4336~A.

\subsection{Ground-based photometry}
\label{sec:ground_based_photometry}
We collected photometric data using four sets of facilities as part of the TESS Follow-up Observing Program (TFOP) Sub-Group 1 (SG1) for Seeing-limited Photometry: SPECULOOS-Southern Observatory \citep[SSO;][]{SPC_Laeti,SPC_Daniel}, TRAPPIST-South \citep[TS;][]{TS_Gillon,TS_Jehin}, Las Cumbres Observatory Global Telescope \citep[LCOGT;][]{Brown_2013}, and ExTrA \citep{Bonfils2015}. The observations are summarized in Table \ref{table_observations}, {and the quality of the light curves is quantified by their RMS value, it is given in Table \ref{table_baseline_error_fit}.} \\
{All but the ExTrA data reduction and analysis was done using a custom pipeline for image processing and photometric extraction built with the \texttt{prose}\footnote{\url{https://github.com/lgrcia/prose}} package \citep{2022_prose,prosesoft}.
First, the image calibration was performed, and the images aligned using \texttt{twirl}\footnote{\url{https://github.com/lgrcia/twirl}}. We detected the stars using an implementation of DAOPhot from Photutils \citep{photutils}, and used \texttt{ballet}\footnote{\url{https://github.com/lgrcia/ballet}} as centroiding algorithm. The photometric extraction was done on a set of 30 circular apertures ranging between 0.5 and twice the value of the Full Width at Half Maximum (FWHM) of the target's Point-Spread Function (PSF), and the position of the background annulus was selected on the basis of the lack of contaminant in the vicinity of the target. We performed differential photometry on all data sets to retrieve our light curves. In doing so we selected comparison stars in the field as well as apertures minimizing the white and red noise in the target light curve as calculated by \cite{2006MNRAS_Pont}. We treated each observation separately, and the number of comparison stars varied between 3 and 9. Given the proximity of stars A and B, the flux in the aperture of the TRAPPIST-South observations is contaminated as it contains fully the PSFs of the two stars. This is taken into account in the global analysis where we fit the dilution factor (see Section \ref{sec:global_analysis}). For the observations with the 1-m class telescopes (SSO and LCO), the stars are completely resolved with a negligible overlap of the PSFs. In that case, we systematically selected TOI-4336~B as a comparison star. }

\begin{table*}
\centering
\caption{Summary of the ground-based follow-up observations obtained for the validation of TOI-4336~A~b.}
\label{table_observations}
\begin{tabular}{lccccccc}
\toprule
\toprule
\vspace{0.15cm}
\bf{Observatory}  & \bf{Filter} & \bf{Date} & \bf{Coverage} & \bf{Exp. time (s)} & \bf{FWHM} (\arcsec) & \bf{Aperture} (\arcsec) & {Measurements} \\
\toprule
\toprule
\vspace{0.12cm}
TRAPPIST-South & \textit{Sloan-z'} & 2021\,Apr\,30  &  Full & 15 & {2.62} & {11.11} & {1001} \\
\vspace{0.1cm}
SSO/Europa & \textit{Sloan-r'} & 2021\,Jun\,18 & Ingress & 10& {1.96} & {3.04} & {647}  \\
\vspace{0.1cm}
SSO/Ganymede & \textit{Sloan-r'} & 2021\,Jun\,18 & Ingress & 10 & {2.31} & {3.44} & {674} \\ 
\vspace{0.1cm}
TRAPPIST-South & \textit{Sloan-z'} & 2021\,Jun\,18  &  Full & 15& {2.42} & {10.69} & {1029}  \\
\vspace{0.1cm}
LCO (CTIO) & \textit{Pan-STARRS-zs} & 2021\,Jun\,18 & Full & 45 & {1.64} & {2.35} & {308} \\
\vspace{0.1cm}
TRAPPIST-South & \textit{Sloan-z'} & 2021\,Aug\,06 & Ingress & 15 & {2.99} & {10.59} & {443}  \\
\vspace{0.1cm}
LCO (SAAO) & \textit{Sloan-g'} & 2022\,Feb\,18 & Ingress & 150  & {1.76} & {2.54} & {54} \\ 
\vspace{0.1cm}
SSO/Europa & \textit{Sloan-g'} & 2022\,Apr\,08 & Full & 24  & {1.19} & {2.38} & {473} \\
\vspace{0.1cm}
LCO (CTIO) & \textit{Sloan-g'} & 2022\,Apr\,08 & Ingress  & 150  & {1.76} & {3.00} & {60} \\
\vspace{0.1cm}
LCO (CTIO)  & \textit{Pan-STARRS-zs} & 2022\,Apr\,08 & Ingress & 45  & {1.53} & {2.30} & {144} \\
\vspace{0.1cm}
SSO/Europa & \textit{Sloan-r'} & 2022\,May\,27 & Full & 10  & {1.11} & {2.60} & {1415}  \\
\vspace{0.1cm}
SSO/Io & \textit{Sloan-g'} & 2022\,May\,27 & Full & 24  & {1.70} & {1.88} & {685} \\
\vspace{0.1cm}
TRAPPIST-South & \textit{Sloan-z'} & 2022\,May\,27 & Full & 15  & {1.82} & {11.91} & {857} \\
\vspace{0.1cm}
ExTrA {(Tel. 2)} & $1.2~\rm \mu m$ & 2022\,May\,27 & Full & 60 & {1.045} & {8.00} & {351} \\
\vspace{0.1cm}
ExTrA {(Tel. 3)} & $1.2~\rm \mu m$ & 2022\,May\,27 & Full & 60 &  {1.367}&  {8.00} & {351} \\
\vspace{0.1cm}
SSO/Io & \textit{Sloan-r'} & 2023\,Mar\,17 & Full & 10 & {1.56} & {2.52} & {1323} \\
\vspace{0.1cm}
SSO/Europa & \textit{Sloan-r'} & 2023\,Mar\,17 & Full & 10 & {1.11} & {2.52} & {1321} \\
\vspace{0.1cm}
LCO (CTIO) & \textit{Pan-STARRS-zs} & 2023\,Mar\,17 & Full & 45 & {1.48} & {2.47} & {207} \\
\vspace{0.1cm}
LCO (CTIO) & \textit{Sloan-g'} & 2023\,Mar\,17 & Full & 150  & {2.71} & {3.28} & {89}  \\
\midrule
\vspace{0.12cm}
{SSO/Callisto} & {\textit{Sloan-r'} }& {2021\,Jul\,13 }& {Monitoring} & {10 } & {2.77} & {2.15} & {829} \\
\vspace{0.1cm}
{SSO/Callisto} & {\textit{Sloan-r'}} & {2021\,Jul\,14} & {Monitoring} & {10 } & {3.32} & {3.14} & {924} \\
\vspace{0.1cm}
{SSO/Callisto }& {\textit{Sloan-r'}} & {2021\,Jul\,15} & {Monitoring} & {10 } & {2.63} & {2.64} & {908} \\
\vspace{0.1cm}
{SSO/Callisto} & {\textit{Sloan-r'}} & {2021\,Jul\,16 } & {Monitoring} & {10}  & {2.51} & {2.97} & {885} \\
\vspace{0.1cm}
{SSO/Callisto} & {\textit{Sloan-r'}} & {2021\,Jul\,17} & {Monitoring} & {10}  & {1.96} & {2.48} & {660} \\
\vspace{0.1cm}
{SSO/Callisto} & {\textit{Sloan-r'}} & {2021\,Jul\,18} & {Monitoring} &{ 10}  & {1.92} & {2.15} & {856} \\
\vspace{0.1cm}
{SSO/Io} & {\textit{Sloan-r'}} & {2022\,Mar\,10} & {Monitoring} & {10}  & {1.39} & {2.74} & {1404} \\
\vspace{0.1cm}
{SSO/Europa} & {\textit{Sloan-r'}} & {2022\,Mar\,11} &{ Monitoring} & {10}  & {1.36} & {2.85} & {1411} \\
\vspace{0.1cm}
{SSO/Europa} & {\textit{Sloan-r'}} & {2022\,Mar\,12} & {Monitoring} & {10}  & {1.36} & {2.06} & {236} \\
\vspace{0.1cm}
{SSO/Io} & {\textit{Sloan-r'}} & {2022\,Mar\,13} & {Monitoring} & {10}  & {1.34} & {3.07} & {767} \\
\vspace{0.1cm}
{SSO/Europa} & {\textit{Sloan-r'}} & {2022\,Mar\,14} & {Monitoring} & {10}  & {1.21} & {2.62} & {783} \\
\vspace{0.1cm}
{SSO/Europa} & {\textit{Sloan-r'}} & {2022\,Mar\,15} & {Monitoring} & {10}  & {1.24} & {2.51} & {792} \\
\vspace{0.12cm}
\end{tabular}
\end{table*}

\subsubsection{TRAPPIST-South}

{TRAPPIST-South is a 0.6m Ritchey-Chrétien telescope located at La Silla Observatory in Chile \citep{TS_Gillon,TS_Jehin}. It is equipped with an FLI ProLine CCD camera with a pixel scale of $0.64''$, providing a field of view of $22' \times 22'$. }
We observed four transits of TOI-4336~A~b with this facility in the \textit{Sloan-z'} filter with 15~s exposures on 2021\,Apr\,30, 2021\,Jun\,18, 2021\,Aug\,6, and 2023\,May\,27.

\subsubsection{SPECULOOS-South}

{The SPECULOOS Southern Observatory, located at ESO Paranal Observatory in Chile, is composed of four 1.0m F/8 Ritchey-Chrétien telescopes \citep{SPC_Laeti,SPC_Daniel}, named after the Galilean moons Io, Europa, Ganymede, and Callisto. All telescopes are equipped with a deep-depletion Andor iKon-L $2k \times 2k$ CCD camera with a total field of view of $12'$ for a pixel scale of $0.35''$ \citep{SPC_Artem}. }
We collected data on the nights of 2021\,Jun\,18, 2022\,Apr\,8, 2022\,May\,27, and 2023\,March\,17 with SSO in the \textit{Sloan-r'} filter with 10s exposures and the \textit{Sloan-g'} filter with 24s exposures. 

\subsubsection{LCOGT}

A total of six transits were obtained with LCOGT  in the \textit{Sloan-g'} with 150s exposures and Pan-STARRS $zs$ with 45s exposures 
on the nights of 2021\,Jun\,18, 2022\,Feb\,18, 2022\,Apr\,8, and 2023\,March\,17. 
{The telescopes are equipped with 4096x4096 SINISTRO Cameras, having an image scale of $0.389\arcsec$ per pixel, resulting a  field of view of 26'x26'.} The raw data were calibrated by the standard LCOGT {\tt BANZAI} pipeline \citep{McCully_2018SPIE10707E} and photometric measurements were extracted using the {\tt prose} package, similarly to the TRAPPIST and SPECULOOS data. 

\subsubsection{ExTrA}
 {The ExTrA facility, located at ESO's La Silla Observatory, is composed of three 0.6m telescopes feeding a near-infrared multi-object spectrograph.} We simultaneously obtained one transit with two ExTrA telescopes \citep{Bonfils2015} on 2022\,May\,27 in a bandpass centered on 1.21$\mu$m with 60~s exposures and the spectrograph's low resolution mode ($R\sim20$) and 8\arcsec diameter aperture fibers. The resulting ExTrA data were analyzed with custom data reduction software, detailed in \cite{Cointepas2021}.

\section{Validation of the planet}
\label{sec:validation}

\subsection{Archival imaging}

\begin{figure*}
	\centering
        \hspace{-1em}
	\includegraphics[scale=0.33]{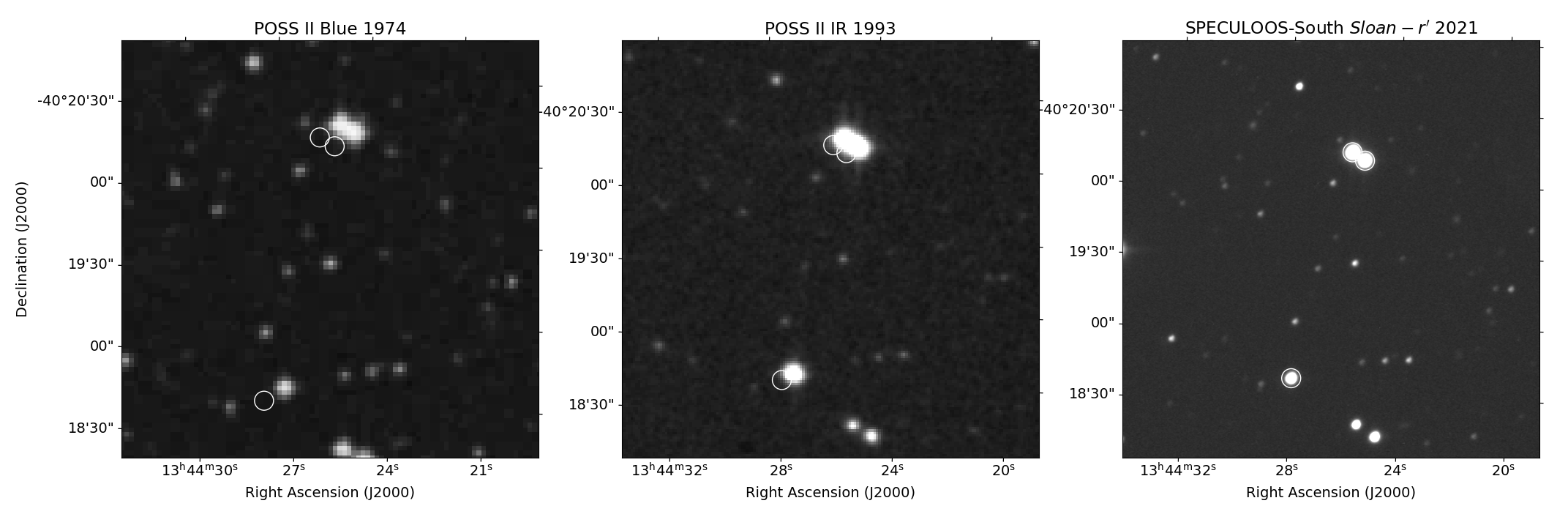}
	\caption{Archival images of the field around the TOI-4336 system. From left to right: 1974 image taken with the blue plate of DSS/POSS-II, 1993 image taken with the InfraRed plate of DSS/POSS-II, and 2021 image taken with SSO in \textit{Sloan-r'}. The {white} circles indicate the position of the stars on the 2021 images.}
	\label{fig:archival_images}
\end{figure*}

Investigating archival imaging of the field of view of the considered target is common practice to look for a possible blend with a background object in the current images. The large proper motion of the TOI-4336 system \citep[166 mas\,yr$^{-1}$;][]{gaiaDR3cat} makes this possible. We inspected DSS/POSS II images in blue and IR \citep{1996DSS_POSS-II,1991_POSSII} spanning 47 years compared to the SSO observation, as shown in Figure \ref{fig:archival_images}. We conclude there is no background star with a limiting magnitude of B$\sim$22.5 at the current position of TOI-4336~A which could potentially be the source of the transit events or {impact our conclusions.}

\subsection{High-angular resolution imaging}


A search for proper motion blend candidates does not eliminate the possibility that the star itself is an unresolved binary. Close-bound stellar companions can confound exoplanet discovery and parameter derivation. If a close companion does exist,  the detected transit signal will yield incorrect stellar and exoplanet parameters \citep{Ciardi2015,2017_Furlan_Howell}. Additionally, the presence of a close companion star {could lead} to the non-detection of small planets residing {within} the same exoplanetary system \citep{Lester2021}. 

The Renormalised Unit Weight Error (RUWE) {quantifies the possible presence of an} unresolved companion or acceleration during the astrometric solution estimation from {\it Gaia\/} measurements \citep{gaiaDR3cat}. As all components of the systems have values above the expected value of $\sim 1.0$ for a single star (see Table \ref{tab:properties_TOI-4336}), {we use high-resolution imaging to constrain the presence of unresolved companions.}

\begin{figure*}
\centering
\includegraphics[scale=0.74]{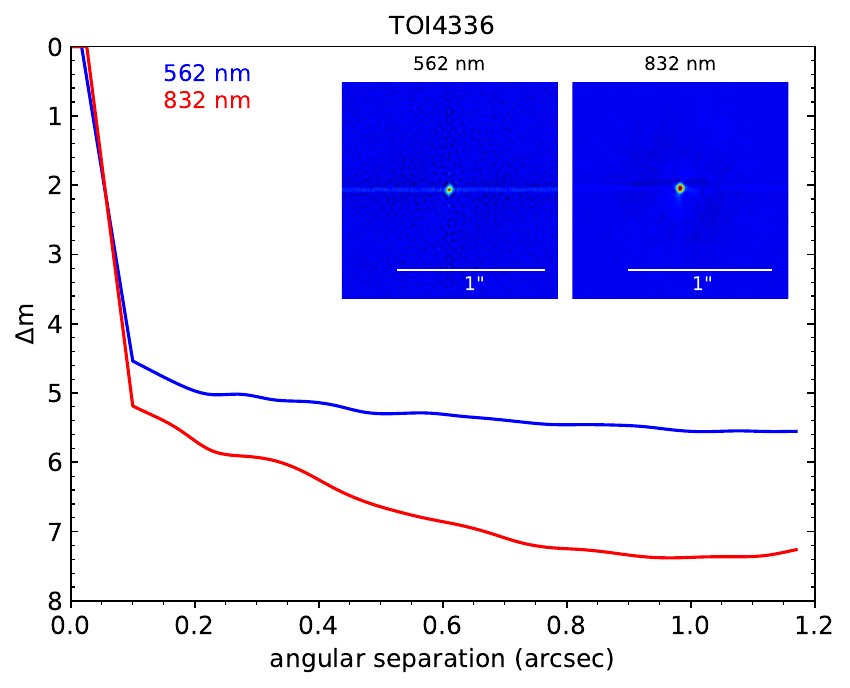}
\caption{Contrast curve obtained with Zorro for TOI-4336~A in two bands (562 nm and 832 nm) and reconstructed speckle image of the observation of 2022\,May\,17. }
\label{Gemini_imaging}
\end{figure*}

TOI-4336~A was observed three times with the Zorro instrument on the Gemini-South 8-m telescope \citep{Scott2021,HF2022}: 2022\,March\,19, 2022\,May\,17, and 2023\,May\,27. Zorro provides simultaneous speckle imaging in two bands (562 nm and 832 nm) with output data products including a reconstructed image with robust contrast limits on companion detections  \citep[e.g.][]{2016Howell}. All three observations were processed with our standard reduction pipeline \citep{Howell2011}, Figure \ref{Gemini_imaging} shows the final contrast curves and the reconstructed speckle images for the 2022\,May\,17 observation. 
TOI-4336~A is an isolated star with no companion brighter than 5-7 magnitudes below that of the target star from the 8-m telescope diffraction limit {of 0.2\arcsec} out to 1.2\arcsec. {This excludes the presence of companion stars with spectral types between M4 and early-L at these angular limits.} At the distance of the TOI-4336 system (d=22.5 pc), {they} correspond to spatial limits of 0.45 to 27\,au.

\begin{figure*}
\begin{minipage}{0.5\textwidth}
\includegraphics[width=\textwidth]{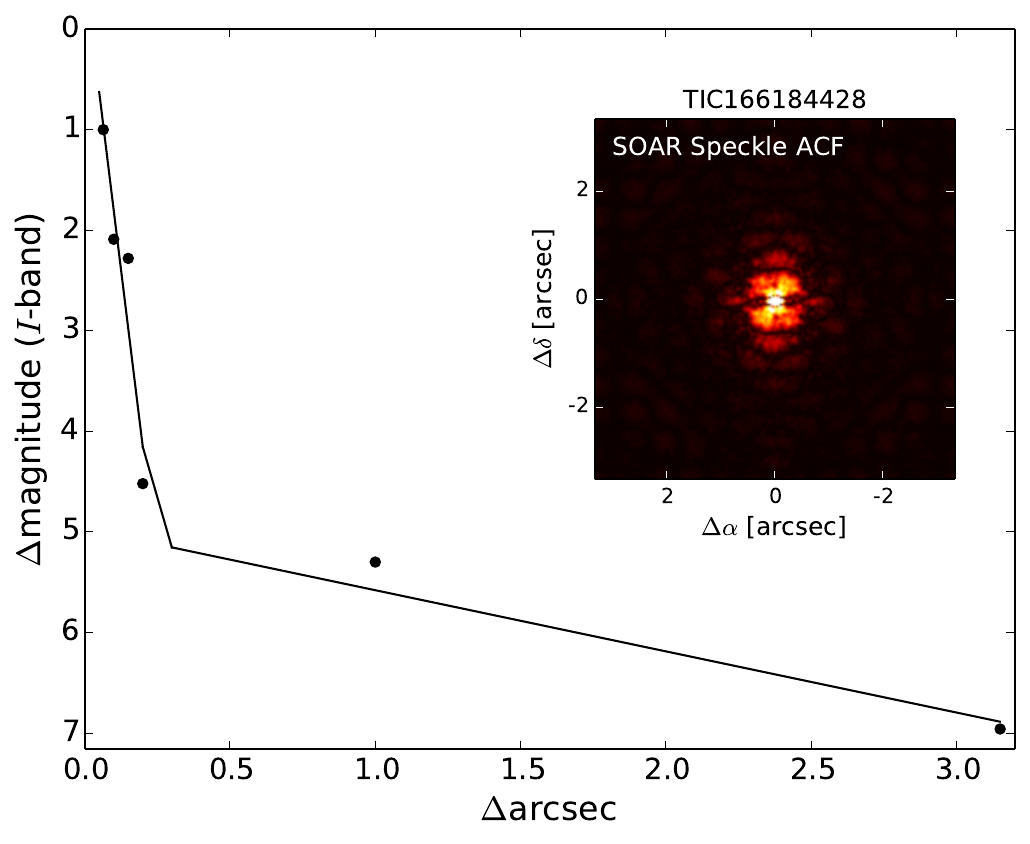}
\end{minipage}
\begin{minipage}{0.5\textwidth}
\includegraphics[width=\textwidth]{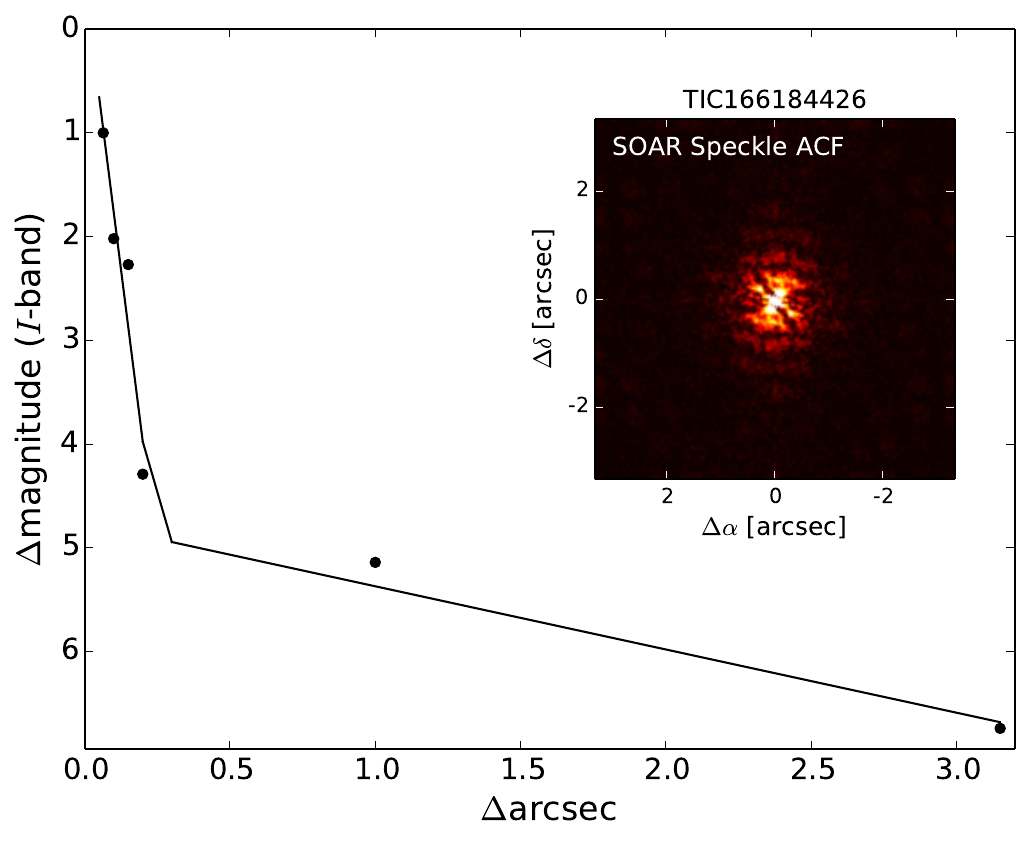}
\end{minipage}
\caption{Speckle autocorrelation functions (ACF) and 5$\sigma$ sensitivity functions of the SOAR observations obtained for A (on the left) and B (on the right). }
\label{SOAR_speckle}
\end{figure*}

We acquired a second set of speckle imaging for the A and B components of the TOI-4336 system with the HRCam instrument of the SOAR telescope \citep{2018_tokovinin_SOAR}, the data were analyzed following the method outlined in \cite{ziegler20}. The observations were obtained on 2021\,Jul\,14, and 2022\,March\,20 for B and A respectively, both in the \textit{Cousins-I} filter, and the contrast curves yielded no companions within 1\arcsec with a contrast of 6.7 and 7 magnitudes. The 5$\sigma$ sensitivity and speckle autocorrelation functions are shown in Figure \ref{SOAR_speckle}.

\subsection{Statistical validation}

We made use of the statistical validation package \texttt{TRICERATOPS}\footnote{\url{https://github.com/stevengiacalone/triceratops}}\citep{triceratops, triceratops_soft} using the same procedures detailed in \cite{Dransfield2023, Pozuelos2023, Barkaoui2023}, to validate the planetary nature of TOI-4336.01. 
The threshold for validation is set at 0.015, and initially we find the False Positive Probability (FPP) to be $0.444$. 
However, upon inspection of the probability breakdown, we found that the Nearby Transiting Planet (NTP) probability on TOI-4336~B is $0.444$, 
while probabilities for all other false positive scenarios are of order $10^{-8}$ or smaller. The reason for the high FPP is therefore that the two stars are blended within the \tess aperture. While we do fold in ground-based observations to the light curve used for the statistical validation to improve the precision, \texttt{TRICERATOPS} only makes use of \tess apertures for the calculation of scenario probabilities.  However, in our ground-based observations, the close binary components of the triple system are resolved and we were able to confirm that the transits are on TOI-4336~A. Therefore, by eliminating this scenario, the FPP for TOI-4336.01 falls to $\sim 10^{-8}$, placing it well below the threshold for validation. 

\section{Global analysis}
\label{sec:global_analysis}
\subsection{Transit analysis}
We performed a global analysis of the photometric data using a custom \texttt{Fortran} code described in \cite{Gillon2012,speculoos2} in which we made use of \texttt{emcee} \citep{emcee}, a Markov Chain Monte Carlo affine invariant ensemble sampler proposed by \cite{2010Goodman_Weare}. We used a combination of quadratic limb-darkening transit model \citep{Mandel2002} and baseline model {to fit the data. The baseline model represents the combination of systematic effects producing correlated (red) noise, such as atmospheric conditions (airmass, full width at half maximum of the point spread function, sky background) or instrumental effects (the variation of the position of the star on the detector along the x and y directions). To remove these effects, we select a linear combination of low-order polynomials with respect to these five parameters, in addition to the time as a parameter, which minimizes the Bayesian Information Criterion  \citep[BIC,][]{schwarz1978}. Fitting simultaneously the transit signal and the correlated noise allows a good propagation of the uncertainties to the derived parameters. The TS light curves are also affected by an additional offset coming from a constraint of the German equatorial mount it is equipped with. As it crosses the meridian, the telescope mount rotates by 180° and the stars fall onto different pixels with varying sensitivity, affecting the flux measurements.} The code also produces $\beta_w$ and $\beta_r$, two scaling factors to apply to the photometric errors to account for an under- or over-estimation of white and red noise in each light curve. The baseline models and the error scaling factors are shown in Table~\ref{table_baseline_error_fit}.


Given that the TS and \tess photometric data sets are contaminated by the second fainter star of the system, we computed the dilution factor as the flux ratio in the TESS band following {the} definition used {in} the code: {$\mathrm{Dil} = \frac{F_{blend}}{F_{source}}$ with F$_{blend}$ the flux of the contaminant and F$_{source}$ the flux of the target star}. {We} used a normal prior with a conservative 3\% error to account for possible faint stars contaminating the apertures. The quadratic limb darkening coefficients $u_1, u_2$ are taken from \cite{Claret_2018_tess} for \tess and the ground-based observations from \cite{Claret_2012_Mdwarfs}, except for the $\mathrm{ExTrA\; 1.2\;\mu m}$ and \textit{Sloan-zs} filters for which the priors were obtained with \texttt{PyLDTK} \citep{pyldtk} and the PHOENIX model atmospheres \citep{phoenix}. We used a conservative value of 0.05 for the uncertainty in the normal prior distributions of these parameters. {We also assumed a normal prior probability distribution function (PDF) for the effective temperature, and the stellar mass and radius based on the results described in Section \ref{sec:sed} (see Table \ref{tab:properties_TOI-4336}).}
{The jump parameters sampled in our analysis are: the effective temperature, the metallicity, the transit epoch, the log of the orbital period, the transit depth as defined by $dF=R_p^2/R_\star^2$, the cosine of the orbital inclination, the log of the stellar density, and the log of the stellar mass. The limb darkening coefficients are also taken as jump parameters following the parametrization of \cite{kippingldcs} for triangular sampling with the quadratic limb darkening law: $q_1 = (u_1 + u_2)^2$ and $q_2 = 0.5u_1(u_1 + u_2)^{-1}$. Finally, we also fitted for the dilution parameters of the TS and \tess data.}
We ran the {\texttt{emcee}} fit using two repeats of 100 walkers with 1000 steps per walker to explore efficiently the full parameter space. We monitored the convergence of the fit using the Gelman-Rubin statistic \citep{1992_gelman_rubin_stat} {which allowed to check that the two independent \texttt{emcee} analyses had produced consistent posterior PDFs for the jump parameters}. Table \ref{tab:param} shows the results of the fit for a fixed depth across all filters. {This includes the jump parameters as well as the derived parameters, for each we give the median of the posterior distribution and the $1-\sigma$ interval. The posterior distributions of the jump parameters are given in the form of a \texttt{corner} plot \citep{corner} in Figure 
 \ref{fig:posteriors}.}

\begin{figure*}
	\centering
	\includegraphics[scale=0.5]{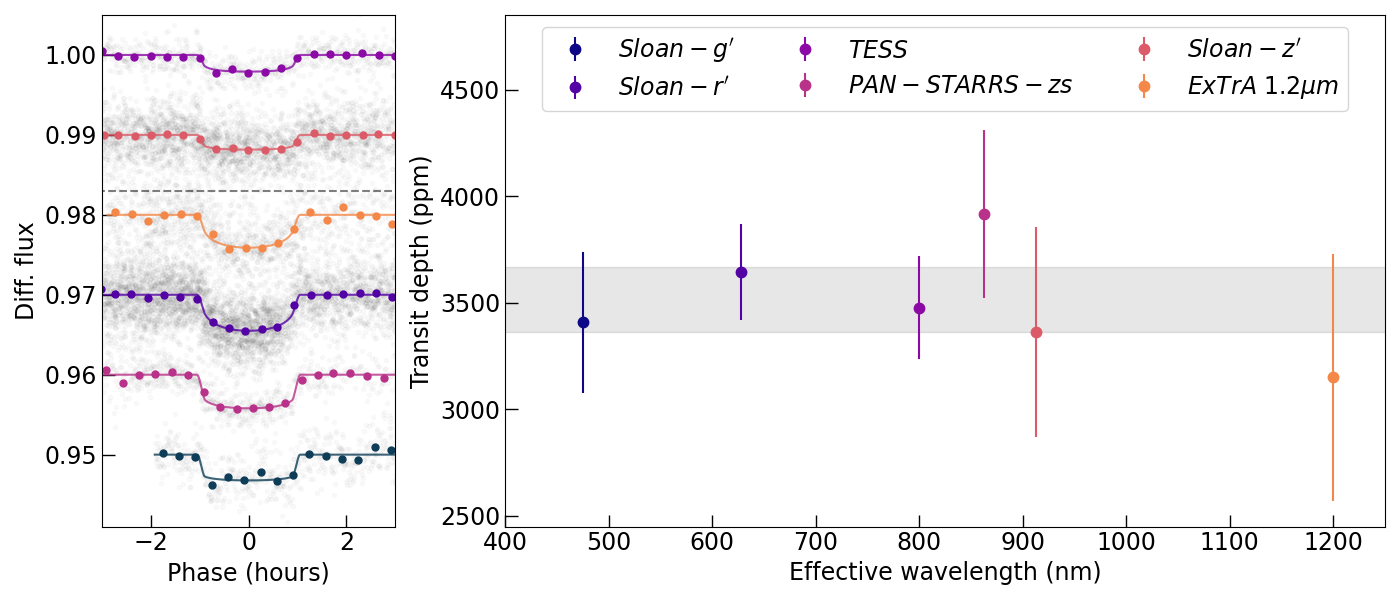}
	\caption{On the left: phase folded transits from TESS and ground-based observations. The filters above the dashed line correspond to the \tess and TRAPPIST-South observations which are diluted by TOI-4336~B. On the right: comparison of the depths obtained for the chromatic fit. All the bands agree within 1-$\sigma$, and the shaded region corresponds to the depth obtained from the achromatic fit for comparison. }
	\label{tess_photometry_sectors}
\end{figure*}

We then performed a chromaticity check, using the same priors but allowing the depth to vary as a function of wavelength. All the depths found agree within 1-$\sigma$, they are shown in Figure \ref{tess_photometry_sectors}. We also performed an eccentric fit, {adding $\sqrt{e_p}$cos$\omega_p$ and $\sqrt{e_p}$sin$\omega_p$ as jump parameters with $e_p$ the eccentricity and $\omega_p$ the argument of periapsis, both of these quantities left as free parameters with no priors. We find a value of $0.12_{-0.09}^{+0.18}$ for the eccentricity.} Following the Bayesian Information Criterion approximation to the Bayes Factor outlined in \cite{wagenmakers_practical_2007_BIC_BF}, we find a Bayes Factor of {10654} which heavily favors the circular fit. {Finally, we performed an analysis allowing the transit timings to vary to check for the existence of Transit Timing Variations (TTVs) which could be indicative of a third body in the system. We found the fitted timings to be consistent with no TTVs down to a few minutes, as shown on Figure \ref{App:TTVs}. The lack of detected TTVs does not allow us to conclude on the presence of an additional planet orbiting TOI-4336~A.}

\begin{table*}[hbt]
   \centering
   \caption{Properties of TOI-4336~A and TOI-4336~A~b based on our global transit model (see Sect. \ref{sec:global_analysis}).}
	\begin{tabular}{lccc}
    	\toprule
    	\toprule
    	{Parameters} & {Values} & {Priors} \\
    	\midrule
    	\midrule
    	\vspace{0.1cm}
    	Luminosity, $L_{\star}$ \small{($\:L_\odot$)} & {$0.0114_{-0.0012}^{+0.0013}$} & - \\
    	\vspace{0.1cm}
             {Metallicity, [Fe/H] (dex)}  & {$-0.209_{-0.013}^{+0.012}$} & {$\mathcal{N}$ (-0.20, 0.12$^2$)} \\
    	\vspace{0.1cm}
    	Effective temperature, $T_{\mathrm{eff}}$ \small{(K)} & {$3298_{-73}^{+75}$} & $\mathcal{N}$ (3300, 75$^2$) \\
    	\vspace{0.1cm}
    	Mass, $M_{\star}$ \small{($M_{\odot}$)} & {$0.331\pm0.015$} & $\mathcal{N}$(0.330, 0.015$^2$) \\
    	\vspace{0.1cm}
    	Radius, $R_{\star}$ \small{($R_{\odot}$)} & {$0.328_{-0.010}^{+0.010}$} & $\mathcal{N}$(0.331, 0.010$^2$) \\
    	\vspace{0.1cm}
    	Density, $\rho_{\star}$ \small{($\rho_{\odot}$)} & {$9.4_{-0.9}^{+1.06}$} & $-$ \\
    	\vspace{0.1cm}
    	Log surface gravity, log $g_{\star}$ \small{(cgs)} & {$4.927_{-0.033}^{+0.034}$} & $-$ \\
    	\vspace{0.1cm}
    	Quadratic limb-darkening coefficient $u_{1,\:g’}$ & {$0.475_{-0.050}^{+0.048}$} & $\mathcal{N}$(0.479, 0.050$^2$) \\
    	\vspace{0.1cm}
    	Quadratic limb-darkening coefficient $u_{2,\:g’}$ & {$0.354_{-0.048}^{+0.049}$} & $\mathcal{N}$(0.349, 0.05$^2$) \\
    	\vspace{0.1cm}
    	Quadratic limb-darkening coefficient $u_{1,\:r’}$ & {$0.551 _{-0.048}^{+0.044}$} & $\mathcal{N}$(0.532, 0.050$^2$) \\
    	\vspace{0.1cm}
    	Quadratic limb-darkening coefficient $u_{2,\:r’}$ & {$0.303_{-0.048}^{+0.049}$} & $\mathcal{N}$({0.278}, 0.050$^2$) \\
    	\vspace{0.1cm}
    	Quadratic limb-darkening coefficient $u_{1,\:z’}$ & {$0.149_{-0.050}^{+0.051}$} & $\mathcal{N}$(0.163, 0.050$^2$) \\
    	\vspace{0.1cm}
    	Quadratic limb-darkening coefficient $u_{2,\:z’}$ & {$0.360 \pm 0.053$} & $\mathcal{N}$(0.368, 0.050$^2$) \\
    	\vspace{0.1cm}
    	Quadratic limb-darkening coefficient $u_{1,\:zs}$ & {$0.284_{-0.047}^{+0.050}$} & $\mathcal{N}$(0.276, 0.050$^2$) \\
    	\vspace{0.1cm}
    	Quadratic limb-darkening coefficient $u_{2,\:zs}$ & {$0.237_{-0.047}^{+0.048}$} & $\mathcal{N}$(0.221, 0.050$^2$) \\
    	\vspace{0.1cm}
    	Quadratic limb-darkening coefficient $u_{1,\:TESS}$ & {$0.165_{-0.050}^{+0.049}$} & $\mathcal{N}$({0.157}, 0.050$^2$) \\
    	\vspace{0.1cm}
    	Quadratic limb-darkening coefficient $u_{2,\:TESS}$ & {$0.455_{-0.050}^{+0.051}$} & $\mathcal{N}$({0.454}, 0.050$^2$) \\
    	\vspace{0.1cm}
    	Quadratic limb-darkening coefficient $u_{1,\:\mathrm{ExTrA 1.2 \mu m}}$ & {$0.180_{-0.053}^{+0.048}$} & $\mathcal{N}$(0.186, 0.050$^2$) \\
    	\vspace{0.1cm}
    	Quadratic limb-darkening coefficient $u_{2,\:\mathrm{ExTrA 1.2 \mu m}}$ & {$0.138_{-0.054}^{+0.050}$} & $\mathcal{N}$(0.146, 0.050$^2$) \\
            Dilution \tess \small{(\%)} & {$86.3 \pm 2.9 $} & $\mathcal{N}$(86, 3$^2$) \\
            Dilution TRAPPIST-South \small{(\%)} & {$86.2_{-2.9}^{+2.7} $} & $\mathcal{N}$(86, 3$^2$) \\
    	\midrule
        \vspace{0.1cm} 
        Transit depth, $\mathrm{d}F$ \small{(ppm)} & {$3487_{-125}^{+126}$}  & $-$ \\
        \vspace{0.1cm}
        Transit impact parameter, $b$ \small{($R_{\star}$)} & {$0.453_{-0.075}^{+0.056}$}  & $-$ \\
        \vspace{0.1cm}
        Orbital period, $P$ \small{(days)} & {$16.336334_{-0.000023}^{+0.000024}$}  & $-$ \\
        \vspace{0.1cm}
        Mid-transit time, \small{$T_{0}$ ($\mathrm{BJD_{TDB}} - 2\:450\:000$)} & {$9335.57319_{-0.00062}^{+0.00068}$}  & $-$ \\
        \vspace{0.1cm}
        Transit duration, $W$ \small{(min)} & {$125.38_{-1.18}^{+1.34}$}  & $-$ \\
        \vspace{0.1cm}
        Orbital inclination, $i_{\mathrm{p}}$ \small{(deg)} & {$ 89.55_{-0.07}^{+0.09}$}  & $-$ \\
        \vspace{0.1cm}
        Orbital semi-major axis, $a_{\mathrm{p}}$ \small{(au)} &  {$0.0872_{-0.0014}^{+0.0013}$}  & $-$ \\
        \vspace{0.1cm}
        Scale parameter, $a_{\mathrm{p}}/R_{\star}$ & {$57.16_{-1.84}^{+2.07}$}  & $-$ \\
        \vspace{0.1cm}
        Radius, $R_{\mathrm{p}}$ \small{($R_{\oplus}$)} & {$2.12_{-0.09}^{+0.08}$} & $-$ \\
        \vspace{0.1cm}
        Stellar irradiation, $S_{\mathrm{p}}$ \small{($S_{\oplus}$)} & {$1.50_{-0.17}^{+0.18}$} & $-$ \\
        \vspace{0.1cm}
        Equilibrium temperature, $T_{\mathrm{eq}}$ \small{(K)} & {$308 \pm$ 9}& $-$ \\

        \end{tabular}
        
\small
\label{tab:param}
\end{table*}

{\subsection{Search for additional candidates}\label{ss:sherlock}}

{Given the high occurrence rates of small planets around M stars \citep{dressing2013_occurence_rate,2020_Hsu_occurence_rate_Mstars}, a system hosting one confirmed planet is likely to host additional ones. As the threshold for a possible TOI detection is set at S/N=$7$ by the \tess pipelines, we then used the \sherlock\footnote{\url{https://github.com/franpoz/SHERLOCK}} package to determine if there could be other candidates in the system that would have been missed \citep{2020_sherlock,sherlockp2}. \sherlock is a community pipeline built on robust and deeply tested astrophysical tools that performs an iterative search for signals on a given star for missions such as \tess. To carry on the task, \sherlock follows the next steps: (1) Downloading and preparing the light curves from online databases using the \lightkurve package. (2) Computing a Lomb-Scargle periodogram \citep{lomb1976,scargle1982,lsp2018} to identify stellar variability, and the field of view plots using \tpfplotter. Other pre-processing steps include correction of stellar variability, and adding a high RMS mask to remove outliers. (3) Following a multi-detrend approach, a bi-weight filter provided by the \wotan\footnote{\url{https://github.com/hippke/wotan}} package \citep{wotan} is applied on the light curves using a range of window sizes. (4) The transit search is performed iteratively on the nominal light curve as well as all the detrended light curves using the \transitleastsquares\footnote{\url{https://github.com/hippke/tls}} package \citep{transitleastsquares}. Once a signal is found above a certain threshold of S/N, it is masked for the next iterations, and this goes on until no more signals with sufficient S/N are found or the search reaches a certain number of iterations. (5) Vetting reports can be created for interesting signals in PDF format, where some metrics are computed and flagged in red when they are considered problematic. (6) A statistical validation can be performed using \triceratops. (7) A Bayesian fit using the Nested Sampling algorithm of \allesfitter\footnote{\url{https://github.com/MNGuenther/allesfitter}} \citep{AllesfitterSoft,AllesfitterPaper}, \texttt{Dynesty}\footnote{\url{https://github.com/joshspeagle/dynesty}} \citep{dynesty,sergey_koposov_2023_dynesty}, can be run for a set of selected signals to refine the system parameters. (8) An observation plan can be created based on the results of the fit for a chosen ground-based observatory.}

{For TOI-4336~A, we performed the transit search on the short-cadence data we extracted with custom apertures for Sectors 11, 38, and 64 of \tess, as described in Section \ref{sec:photo_data_TESS}. We first corrected the light curves for dilution using the value of 86.3\% we obtained in the global analysis. We selected a range of ten window sizes between 0.2 and 1.3 days to generate the detrended light curves. We explored periods between 0.5 and 20 days, and selected a threshold signal of S/N$\geq$5, with a maximum number of five runs. We ran the transit search algorithm considering all three sectors combined in addition to individual sectors. We found the signal of TOI-4336~A b to be recovered easily in the first run of all our analyses with an S/N of 18.70. We also found a second potential candidate in the combined search with a period of $7.59$ d,  a duration of $1.74$ h, and a depth of $1.18$ ppt. We denote it as TIC 166184428.02\footnote{The candidate has been submitted as a Community TESS Object of Interest \url{https://exofop.ipac.caltech.edu/tess/target.php?id=166184428}}, assuming it corresponds to a transiting planet orbiting TOI-4336 A as well. The candidate is recovered in four out of the ten detrended light curves in the second transit search run with an S/N of 5.35. We performed the vetting stage and TIC 166184428.02 passed all the tests except for two: (1) The transit source offset computed from difference imaging, a method introduced in \cite{bryson2017}. (2) The per-pixel Box Least Squares \citep[BLS,][]{blspaper} search showed some deviation from the target. This method uses a fixed epoch and period for the BLS run on each pixels and computes the S/N of the detections. The S/N are then normalized and the centroid is found, then compared to the one found by difference imaging. These metrics are affected by the significant contamination from nearby sources which shifts the centroid on the Target Pixel Files. Given the presence of TOI-4336 B in the aperture, we do not consider these failed tests to be critical. Finally, we performed the fitting stage of \sherlock to recover the system parameters. The results of the Nested Sampling fit are found in Table \ref{tab:allesfit.02}, the folded light curve and the posterior distributions of the fitted parameters can be found in Figures \ref{fig:allesfit_lc} and \ref{fig:allesfit_corner}. \allesfitter uses the parameters found in the transit search as uniform priors for the Nested Sampling fit and a GP with a Matern 3/2 kernel for detrending.}

{TIC 166184428.02 is at the limit of the detection threshold we set in our transit search with \sherlock. The shallow depth of the signal makes it challenging to check whether the single events look consistent in shape with the transit of a planet. We also performed an independent transit search on the \tess data to assess the detection limits of the data, and we did not recover the candidate (see Section \ref{ss:injrec}). }

{\subsection{Detection limits and injection-recovery tests}\label{ss:injrec}}
\begin{figure}
	\centering
	\includegraphics[scale=0.5]{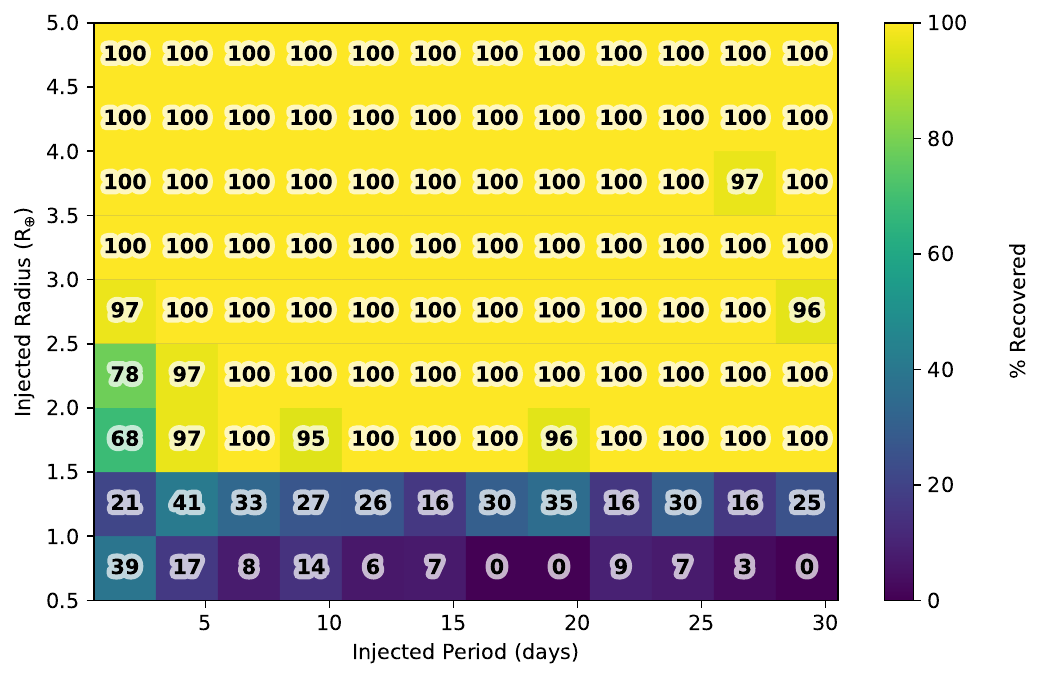}
	\caption{{Results of the injection-recovery tests on TESS light curves, described in Section \ref{ss:injrec}. Almost all planets larger than 1.5R$_{\oplus}$ were recovered successfully up to 30\,d.}}
	\label{fig:tess_injrec}
\end{figure}


{We used the SPECULOOS-Southern Observatory to collect ground-based data to check if monitoring of TOI-4336~A from 1-m class telescopes would allow the detection of smaller planets thanks to an improved precision compared to \tess, as it has been successfully done in \cite{speculoos2}. We gathered twelve nights of photometric observations without any transits, in the \textit{Sloan-r'} filter, the details are shown in Table \ref{table_observations}. The photometric light curves for each night were extracted using the \texttt{prose} package, following the method described in Section \ref{sec:ground_based_photometry} as for other ground-based data, and combined. We ran transit-search pipeline \texttt{occultence}\footnote{\url{https://github.com/catrionamurray/occultence}} to search for additional transiting planets around TOI-4336~A with both SPECULOOS and \tess (Section \ref{sec:photo_data_TESS}). \texttt{occultence} consists of several steps: cleaning (cosmic ray and spurious data point removal), an initial search to mask transit structures (for detrending) using Box Least Squares \citep[BLS,][]{blspaper}, a detrending method, followed by a final transit search on the detrended light curve using BLS. For SPECULOOS detrending, we chose ridge regression \citep{1970HoerlKennard} to fit polynomials of airmass, full width half maximum (FWHM), sky background, $\delta$x and $\delta$y (change in the position of the target on the CCD). For each night, we fit for all combinations of these parameters with different polynomial orders (up to cubic) and selected the combination with the lowest AIC \citep[Akaike Information Criterion,][]{akaike1973second}, which assesses a model's likelihood of describing the data while penalising larger numbers of parameters. For \tess's detrending step instead we performed GP-detrending to capture remaining correlated noise. We did not detect any transit structures in either SPECULOOS or TESS detrended light curve with SNR$>$3.}

{To assess the detection efficiency of our transit-search pipeline we ran injection-recovery tests on both the SPECULOOS and \tess light curves (with real transits masked). For each instrument, we generated transits for 3000 artificial planets using \textsc{PyTransit} \citep{Parviainen2015PYTRANSIT:PYTHON} and injected each in turn into TOI-4336~A's light curves. We used the $Sloan-r'$ and \tess limb-darkening coefficients from Table \ref{tab:param} for SPECULOOS and \tess, respectively. Each planet's remaining parameters (radius $R_{\rm{p}}$, period $P$, and inclination $i$) were drawn from the following uniform distributions: $R_{\rm{p}} \sim U(0.5,5.0)\,R_{\oplus}$, $\cos{i} \sim U(\cos{i_{\rm{min}}},\cos{i_{\rm{max}}})$, $\log{P} \sim U(\log{0.5},\log{30.5})$\,d, where $U(a,b)$ represents a value drawn from a uniform distribution between $a$ and $b$. $i_{\rm{min}}$ and $i_{\rm{max}}$ are the minimum and maximum inclinations for a transiting planet. Due to the shorter baseline of SPECULOOS data, we limited the explored period range up to 10.5d. The host mass and radius were taken from Table \ref{tab:properties_TOI-4336}. However, the inclination limits depend on the orbital period; therefore, when drawing the planetary parameters, we drew each inclination from a range set by the period. Only circular orbits we considered. The time at which the first transit was injected was also drawn from a uniform distribution, $\phi \sim U(0,1)$, where $\phi$ is the phase of the period, such that the first transit was injected at $\phi P$ from the start of observations.}

{We used the transit-search method described at the start of this section to search for the injected planets. A planet was recovered if at least one epoch from the highest likelihood period from BLS was within 1 hour of an injected transit and the SNR of that transit is $>$3. This recovery criterion allows us to detect even single transits. The results from our injection-recovery for TOI-4336~A with \tess are shown in Figure \ref{fig:tess_injrec}. Due to the day-night cycle and the complications in dealing with ground-based systematics, the injection-recovery results for \tess exceed SPECULOOS for all periods $>1$\,d, though the results are comparable for small ($<$1.5R$_{\oplus}$) planets and very short periods. From the \tess results it is unlikely that there exist additional planets orbiting TOI-4336~A with R$>$1.5R$_{\oplus}$ and P$<$30\,d, though we cannot rule out smaller, or longer period, planets.}

\section{{Discussion}}
\label{sec:discussion}


\subsection{Habitability}
The stellar irradiation from its host, $1.50_{-0.17}^{+0.18}$ $S_{\oplus}$, puts TOI-4336~A~b very close 
at a distance consistent with the inner edge of the empirical HZ of the system (about 1.488 $S_{\oplus}$) \citep{kopparapu2013}, as shown in Figure \ref{fig:TSM_plot}. {The limits of the HZ and the environments of planets orbiting close to these limits are not fully understood. Considering the error bars on the stellar irradiation, the planet could receive less or more irradiation than the irradiation at the empirical HZ limit. Probing the atmosphere of planets at the limit of the HZ will provide further insights into the environments at these stellar irradiations.}  
{A planetary radius of $2.12$ R$_\oplus$ places the planet beyond the radius valley \citep[e.g.][]{2022_gupta_radius_valley_lowmass}, in the realm of mini-Neptunes where planets are likely to have an extended gaseous atmosphere. This expected atmosphere should allow for ease of atmospheric characterization. 
This makes TOI-4336~A~b an interesting target to explore the environment at and around the empirical inner edge limit of the HZ and assess whether it is the likely case of a Mini-Neptune or the less likely case of a rocky planet at the inner edge of the HZ.}

TOI-4336~A~b orbits a host star in a triple system. While multiple host stars can influence the HZ boundaries \citep{kaltenegger2013, Haghighipour2013,2013_Kane_Hinkel_HZ_binary}, the two other {mid-M} stars in the system orbit far enough apart to have no significant influence on the HZ limits. {Using the bolometric luminosity from the SED fit (see Section \ref{sec:sed}) and for a median semi-major axis of 133 au (see Section \ref{sec:orbits}), the added flux from the second star would only add $5.767 10^{-7}$ $S_\oplus$ to the overall irradiation.}
Still, TOI-4336~A~b provides a very interesting, as well as accessible, target to explore the region around the empirical HZ limits. 


\subsection{Formation in the triple system}
\label{sec:formation in triple system}

%

Hundreds of planets have been discovered in binary systems~\citep[e.g.][]{raghavan10,Matson2018}.  Yet binaries -- especially on close or eccentric orbits -- shrink the orbital real estate available for planets~\citep{holman99}.  In addition, an inclined binary can trigger Kozai oscillations~\citep{takeda08}, although this is impeded during the disk phase~\citep{batygin11} and in multiple-planet systems~\citep{innanen97,kaib11}.  In general, wider binaries have a progressively weaker influence.  There is an observed deficit of exoplanets in systems with binaries closer than $\sim 50$~au~\citep{kraus16}. Yet, in some cases, binaries wider than $\sim 1000$~au can reach very eccentric orbits due to external Galactic perturbations and destabilize the orbits of gas giants orbiting at Jupiter- to Saturn-like distances~\citep{kaib13}.  However, there is no evidence that the overall occurrence rate of exoplanets is strongly affected by wide binary companions~\citep[e.g.][]{kraus16,ziegler20} .

For the case of TOI-4336~A~b, we do not expect the triple stellar configuration to have played a strong role in the planet's formation or evolution.  The maximum orbital radius that remains stable in the face of perturbations from a binary is a function of the companion's mass and the binary orbital parameters~\citep{holman99,pilat-lohinger13}. The closer stellar companion of TOI-4336~A has a semi-major axis of 133\,au (see Section~\ref{sec:orbits}). Assuming a binary orbital eccentricity of 0.7 (the median for a thermal distribution) and equal stellar masses, the outermost stable orbit around TOI-4336~A would be about 3\% of the binary semi-major axis, or $\sim4$\,au~\citep[assuming coplanarity between the stellar and planetary orbits;][]{holman99}, more than an order of magnitude larger than the orbital radius of 0.09\,au of TOI-4336~A~b. { If the binary's periastron distance were 5 au, which is the minimum value in the 1-$\sigma$ contours from the analysis in Section \ref{sec:orbits}, then the binary's eccentricity would have been so high that it could potentially have disrupted planet formation entirely.  Given the existence of the planet, we expect that the binary's eccentricity is likely no higher than 0.7-0.8, to avoid any drastic consequences during planetary growth.}
The second companion, with a semi-major axis of 2915\,au, would have a much smaller effect. It therefore seems unlikely that the companion stars played a direct role in the formation or evolution of the sub-Neptune TOI-4336~A~b.  


\subsection{Prospects for detailed characterization}
\label{sec:characterization}

We explore the possibility to constrain the planet mass using high-resolution spectroscopy. Following the mass--radius relationship by \cite{2017ApJ...834...17C}, we estimate the planet mass to be $5.4^{+4.1}_{-2.2}\,{\rm M_{\oplus}}$. Measuring the implied semi-amplitude of the spectrocopic orbit, $2.9\pm1.6\,{\rm m\,s^{-1}}$, would require high-precision spectroscopy using stabilised spectrographs such as ESPRESSO \citep{pepe10}. 
{Thanks to its apparent brightness (Vmag$\sim$12.9), the star is within reach of ESPRESSO with a $3.2\pm1.8\,{\rm \sigma}$ detection of the planet with 15 spectra.} 
Because of its infrared brightness (Hmag=8.9), we expect detection to be also possible using NIRPS \citep{Bouchy17}, which has been proven to show a long-term stability of $2-3\,{\rm m\,s^{-1}}$ for bright targets. There is already an ongoing radial velocity campaign with both these facilities to measure the mass of TOI-4336~A~b.

Given the infrared brightness of the host (Kmag $\sim$ 8.6) and the favourable planet-to-star radius ratio due to the low-mass host star, we assess the suitability of TOI-4336~A~b for characterization of an atmosphere. We first calculated the transmission spectroscopy metric \citep[TSM,][]{kempton2018} as it is a convenient metric to compare the amenability for characterization of different planets. We find that TOI-4336~A~b has an exceptionally high TSM of {83$_{-4}^{+5}$ in the temperate sub-Neptune category}, making it even more favourable than some of the best-studied sub-Neptunes for a detailed atmospheric characterization by transit transmission spectroscopy with \hst and \jwst \citep[e.g. LHS~1140~b,][]{Belly2021}. 

\begin{figure*}
    \centering
    \includegraphics[scale=0.6]{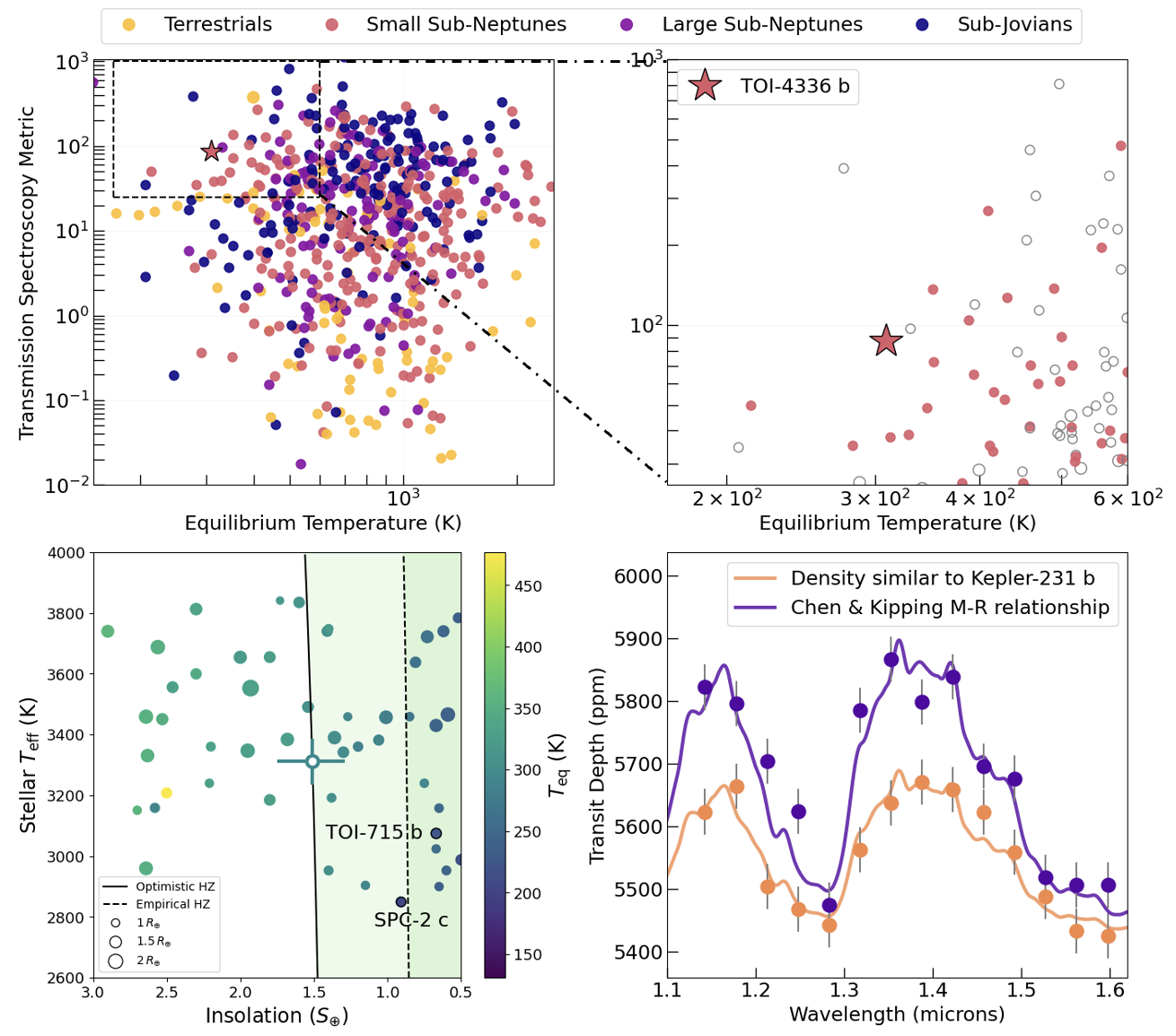}
    \caption{{\it Top left panel:} Complete sample of all known exoplanets with measured masses (Data from NASA Exoplanet Archive, {Apr 2024}, https://exoplanetarchive.ipac.caltech.edu/). {Following \cite{kempton2018}, terrestrial planets correspond to planetary radii below 1.25$\rm{R}_{\oplus}$, small sub-Neptunes radii between 1.25 and 2.75$\rm{R}_{\oplus}$, large sub-Neptunes radii between 2.75 and 4$\rm{R}_{\oplus}$, and sub-Jovians are planets with radii between 4 and 10$\rm{R}_{\oplus}$}. TOI-4336~A~b is shown by the star symbol. {\it Top right panel:} Zoomed in view that puts TOI-4336~A~b in the context of small sub-Neptunes only. 
    {\it Bottom left panel:} Stellar effective temperature as a function of insolation of transiting exoplanets orbiting hosts cooler than 4000~K. The solid black line denotes the empirical HZ boundary and the dashed line the conservative one \citep{kopparapu2013}. The size of points scales with the planetary radius. The points are coloured according to their equilibrium temperature. TOI-4336~A~b is highlighted with errorbars, and it is well placed at the inner edge of the HZ of its host star.
    {\it Bottom right panel:}  Synthetic transmission spectra for TOI-4336~A~b, assuming a typical H/He-rich atmosphere with isothermal temperature profile at the planet's equilibrium temperature ($\sim300\rm K$). The simulated data points are for the Wide Field Camera 3 instrument of \hst. 
    }
    \label{fig:TSM_plot}
\end{figure*}

We then used \texttt{ExoTransmit} \citep{exotransmit} to calculate two simulated transmission spectra for TOI-4336~A~b, assuming a typical H/He-rich atmosphere with {an} isothermal {P-T} profile at the planet's equilibrium temperature ($\sim300\rm K$). {We also assume equilibrium chemistry and include all the opacities available in \texttt{ExoTransmit}. The spectra are shown in the bottom right panel of Figure \ref{fig:TSM_plot}.} The first spectrum assumes a mass equal to the estimate derived from empirical mass-radius relations presented in \cite{2017ApJ...834...17C}, while the second allows for a case where the planet has a similar density to one of the densest known mini-Neptune \citep[Kepler-231~b,][]{Hadden2014}. A lower mass would simply increase the scale height making atmospheric investigations easier. We find that for the mass estimate of $\sim5.4\,{\rm M_{\oplus}}$, we could detect atmospheric features with $>16\sigma$ significance {in the case of a transparent atmosphere} with just three transits of \hst. Even in the higher mass limit of $\sim10.1\,{\rm M_{\oplus}}$, our simulations show that we could still detect a putative atmosphere to $>5\sigma$ with the same observing setup. This makes TOI-4336~A~b one of the most promising sub-Neptunes for atmospheric investigations. {However, the presence of clouds and/or hazes could hinder the detection of atmospheric signals. Atmospheric exploration with \hst is then crucial to determine whether this is the case or not}. In fact, a proposal for these observations has already been accepted {for} \hst and the campaign to measure the atmosphere of this exquisite sub-Neptune is underway. We expect this planet to become one of the best studied in its category in the coming years.

\section{Conclusions}
\label{sec:conclusions}
{Our analyses show that TOI-4336~A~b is a temperate sub-Neptune orbiting a nearby star part of a triple M-dwarf system. We tested the chromaticity of the transit with multi-color photometry and found the transit depths to be consistent within 1-$\sigma$ in all the bands. We found no significant variations in the transit depth between odd and even transits, which could be an indicator of binarity or a blended source. In addition, we statistically validate the planetary nature using archival imaging spanning 47 years of observations, and high angular resolution imaging, which excludes unresolved companions with spectral type between M4 and early-L between 0.2 and 1.2\arcsec\ in separation. Our global model yields a planetary radius of $2.12_{-0.09}^{+0.08}$ R$_\oplus$ and an orbital period of $16.33$ days, resulting in an incident irradiation of $1.50_{-0.17}^{+0.18}$ S$_{\oplus}$. This incident irradiation would lead to an equilibrium temperature estimate of {$308 \pm 9$} K, assuming a Bond albedo of zero and perfect heat redistribution ($f=1/4$). We find the host to be an {M3.5} star ($T_{\mathrm{eff}}$ = {$3298_{-73}^{+75}$}\,K, {$M_\star$ = $0.33 \pm 0.02$ }$M_{\odot}$), with estimated semi-major axes of 133\,au and 2915\,au with respect to the other two {mid-M} stars of the triple system.}

{The radius of this new planet puts it most likely in the mini-Neptune category, and is thus likely to have retained an extended atmosphere. The incident radiation places TOI-4336~A b at the inner edge of the empirical HZ, which makes it a good candidate to explore this region and its consequences on habitability. We investigated the implications of the triple star system on the formation of the planet and found that the eccentricity of the closer pair should be no higher than 0.8. In addition, the orbital configuration of the system implies TOI-4336~B did not have any effect on the formation of TOI-4336~A b. Indeed, the planet would need to be at more than 40 times the orbital distance from its host star for the companion to disrupt the formation of the planet. }

{TOI-4336~A b shows similar properties to the widely studied K2-18 b \citep[e.g.][]{2017_K2-18b_Spitzer,2017_K2-18b_HARPS,Benneke2019,2019_K2-18b_Tsiaras}. We examined the suitability of this temperate sub-Neptune for detailed characterization and found it is a prime target for atmospheric studies with \hst and \jwst. The determination of its mass is a key ingredient for the interpretation of transmission spectra. We find both ESPRESSO and NIRPS are appropriate to measure the radial velocities of the planet and constrain its mass. } 



\begin{acknowledgements}
Funding for the \tess mission is provided by NASA's Science Mission Directorate. We acknowledge the use of public \tess data from pipelines at the \tess Science Office and at the \tess Science Processing Operations Center. This research has made use of the Exoplanet Follow-up Observation Program website, which is operated by the California Institute of Technology, under contract with the National Aeronautics and Space Administration under the Exoplanet Exploration Program. This paper includes data collected by the \tess mission that are publicly available from the Mikulski Archive for Space Telescopes (MAST).
Based on data collected by the SPECULOOS-South Observatory at the ESO Paranal Observatory in Chile.The ULiege's contribution to SPECULOOS has received funding from the European Research Council under the European Union's Seventh Framework Programme (FP/2007-2013) (grant Agreement n$^\circ$ 336480/SPECULOOS), from the Balzan Prize and Francqui Foundations, from the Belgian Scientific Research Foundation (F.R.S.-FNRS; grant n$^\circ$ T.0109.20), from the University of Liege, and from the ARC grant for Concerted Research Actions financed by the Wallonia-Brussels Federation. This work is supported by a grant from the Simons Foundation (PI Queloz, grant number 327127).
This research is in part funded by the European Union's Horizon 2020 research and innovation programme (grants agreements n$^{\circ}$ 803193/BEBOP), and from the Science and Technology Facilities Council (STFC; grant n$^\circ$ ST/S00193X/1, and ST/W000385/1).
The material is based upon work supported by NASA under award number 80GSFC21M0002.
Based on data collected by the TRAPPIST-South telescope at the ESO La Silla Observatory. TRAPPIST is funded by the Belgian Fund for Scientific Research (Fond National de la Recherche Scientifique, FNRS) under the grant PDR T.0120.21, with the participation of the Swiss National Science Fundation (SNF). 
Based on data collected under the ExTrA project at the ESO La Silla Paranal Observatory. ExTrA is a project of Institut de Planétologie et d'Astrophysique de Grenoble (IPAG/CNRS/UGA), funded by the European Research Council under the ERC Grant Agreement n. 337591-ExTrA. This work has been supported by a grant from Labex OSUG@2020 (Investissements d'avenir -- ANR10 LABX56).
This work has been carried out within the framework of the NCCR PlanetS supported by the Swiss National Science Foundation.
The Digitized Sky Surveys were produced at the Space Telescope Science Institute under U.S. Government grant NAG W-2166. The images of these surveys are based on photographic data obtained using the Oschin Schmidt Telescope on Palomar Mountain and the UK Schmidt Telescope. The plates were processed into the present compressed digital form with the permission of these institutions.
This work makes use of observations from the LCOGT network. Part of the LCOGT telescope time was granted by NOIRLab through the Mid-Scale Innovations Program (MSIP). MSIP is funded by NSF.
Based in part on observations obtained at the Southern Astrophysical
Research (SOAR) telescope, which is a joint project of the
Minist\'{e}rio da Ci\^{e}ncia, Tecnologia e Inova\c{c}\~{o}es
(MCTI/LNA) do Brasil, the US National Science FoundationÕs NOIRLab,
the University of North Carolina at Chapel Hill (UNC), and Michigan
State University (MSU).  IRAF was distributed by the National Optical
Astronomy Observatory, which was managed by the Association of
Universities for Research in Astronomy (AURA) under a cooperative
agreement with the National Science Foundation.
Some of the observations in the paper made use of the High-Resolution Imaging instrument Zorro obtained under Gemini LLP Proposal Number: GN/S-2021A-LP-105. Zorro was funded by the NASA Exoplanet Exploration Program and built at the NASA Ames Research Center by Steve B. Howell, Nic Scott, Elliott P. Horch, and Emmett Quigley. Zorro was mounted on the Gemini South telescope of the international Gemini Observatory, a program of NSF’s OIR Lab, which is managed by the Association of Universities for Research in Astronomy (AURA) under a cooperative agreement with the National Science Foundation. on behalf of the Gemini partnership: the National Science Foundation (United States), National Research Council (Canada), Agencia Nacional de Investigación y Desarrollo (Chile), Ministerio de Ciencia, Tecnología e Innovación (Argentina), Ministério da Ciência, Tecnologia, Inovações e Comunicações (Brazil), and Korea Astronomy and Space Science Institute (Republic of Korea).
The postdoctoral fellowship of KB is funded by F.R.S.-FNRS grant T.0109.20 and by the Francqui Foundation. This publication benefits from the support of the French Community of Belgium in the context of the FRIA Doctoral Grant awarded to M.T. E.D. acknowledges support from the innovation and research Horizon 2020 program in the context of the  Marie Sklodowska-Curie subvention 945298. BVR thanks the Heising-Simons Foundation for support. MG is F.R.S.-FBRS Research Director. F.J.P acknowledges financial support from the grant CEX2021-001131-S funded by MCIN/AEI/10.13039/501100011033 and through projects PID2019-109522GB-C52 and PID2022-137241NB-C43. KAC and SNQ acknowledge support from the TESS mission via subaward s3449 from MIT. B.-O. D. acknowledges support from the Swiss State Secretariat for Education, Research and Innovation (SERI) under contract number MB22.00046. E. J. is a Belgian FNRS Senior Research Associate. YGMC acknowledges support from UNAM-PAPIIT-IG101321. L.D. is an F.R.S.-FNRS Postdoctoral Researcher. SNR acknowledges support from the French Programme National de Planétologie (PNP).

This research made use of Lightkurve, a Python package for Kepler and TESS data analysis (Lightkurve Collaboration, 2018). This research has made use of the NASA Exoplanet Archive, which is operated by the California Institute of Technology, under contract with the National Aeronautics and Space Administration under the Exoplanet Exploration Program. Resources supporting this work were provided by the NASA High-End Computing (HEC) Program through the NASA Advanced Supercomputing (NAS) Division at Ames Research Center for the production of the SPOC data products. This work made use of \texttt{tpfplotter} by J. Lillo-Box (publicly available in www.github.com/jlillo/tpfplotter), which also made use of the python packages \texttt{astropy}, \texttt{lightkurve}, \texttt{matplotlib} and \texttt{numpy}.


This work has made use of data from the European Space Agency (ESA) mission {\it Gaia\/} (\href{url}{https://www.cosmos.esa.int/gaia}), processed by the {\it Gaia\/} Data Processing and Analysis Consortium (DPAC, \href{url}{https://www.cosmos.esa.int/web/gaia/dpac/consortium}). Funding for the DPAC has been provided by national institutions, in particular the institutions participating in the {\it Gaia\/} Multilateral Agreement. This research has made use of the Washington Double Star Catalog maintained at the U.S. Naval Observatory. This publication makes use of data products from the Two Micron All Sky Survey, which is a joint project of the University of Massachusetts and the Infrared Processing and Analysis Center/California Institute of Technology, funded by the National Aeronautics and Space Administration and the National Science Foundation.

\end{acknowledgements}

\bibliographystyle{aa}
\bibliography{references}


\begin{appendix}

\section{Color-magnitude diagram}
\begin{figure*}[h!]
    \centering
    \includegraphics[width=\columnwidth]{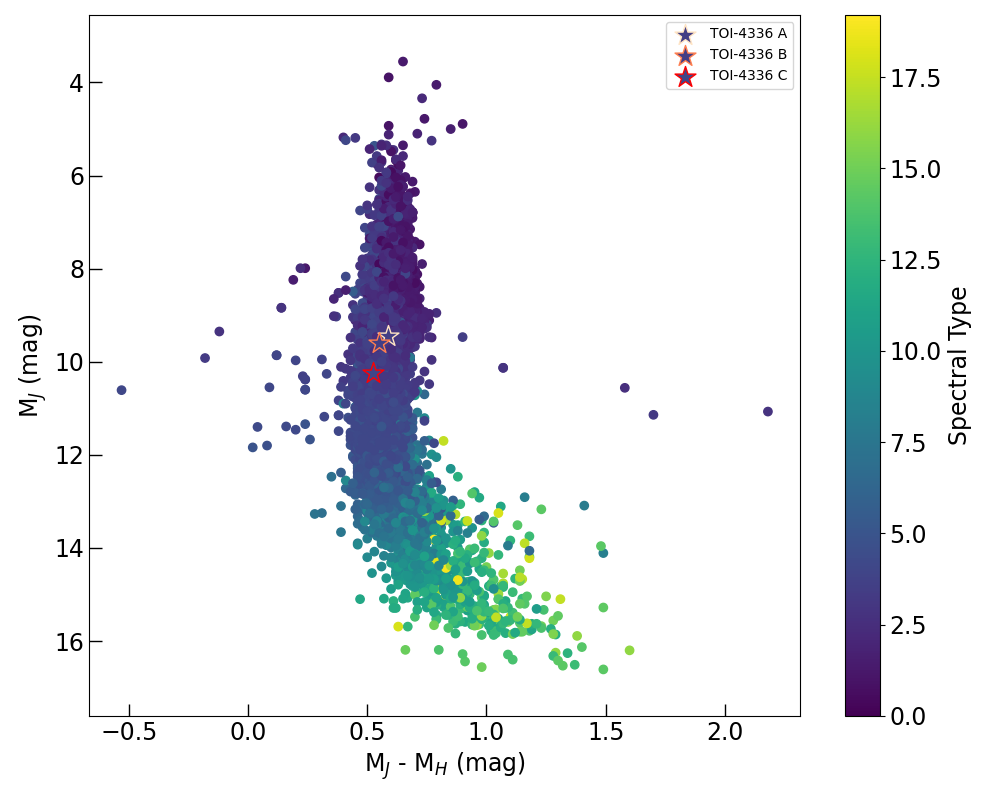}
    \caption{{Color-magnitude diagram for nearby M dwarfs.The TOI-4336 system is marked with a star symbol and color-coded to distinguish the three stars. TOI-4336~C appear redder than A and B, which is consistent with a later spectral type (M4$\pm$0.5 compared to M3.5$\pm$0.5. We used the extended target list of the SPECULOOS survey which gathers over 14 000 M-dwarf stars within 40 parsecs to generate the diagram \citep{SPC_Daniel}.} }
    \label{fig:color-mag}
\end{figure*}

\section{Age estimation}
\begin{figure*}[h!]
    \centering
    \includegraphics[width=\columnwidth]{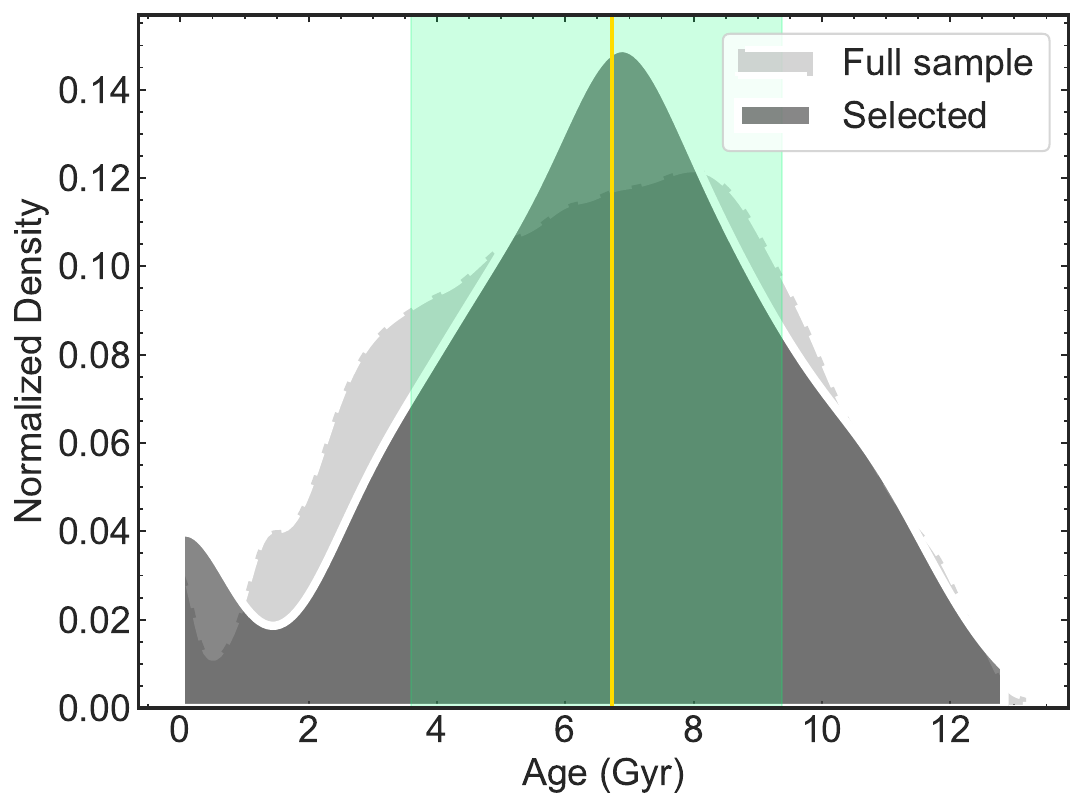}
    \caption{Age distribution the \textit{GALAH} survey (light grey) and the distribution of stars within our selection criteria (dark grey). The yellow line indicates the median age and the yellow bands are the 16~percentile and 84~percentile regions. We estimate an age of  {6.7$_{-3.1}^{+2.7}$~Gyr} for the system. }
    \label{fig:age_estimation}
\end{figure*}



\section{Triple system orbital analysis}
\label{appendix:orbital_fit}

\begin{figure*}[hbt]
    \begin{minipage}[hbt!]{0.5\textwidth}
    \hspace{0.5cm}
        \includegraphics[width=0.95\textwidth]{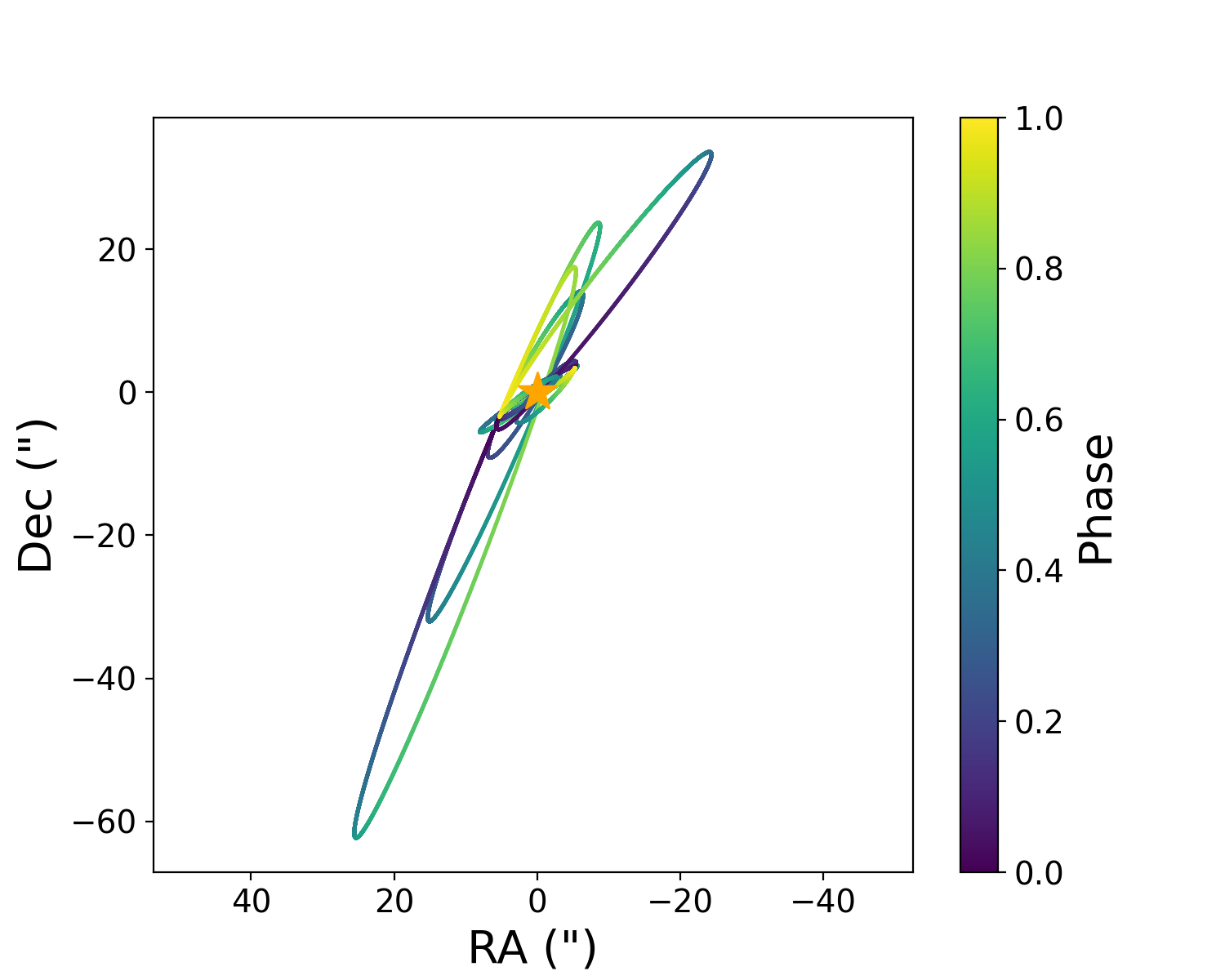}
	\label{AB_orbit}
	\includegraphics[width=0.95
\textwidth]{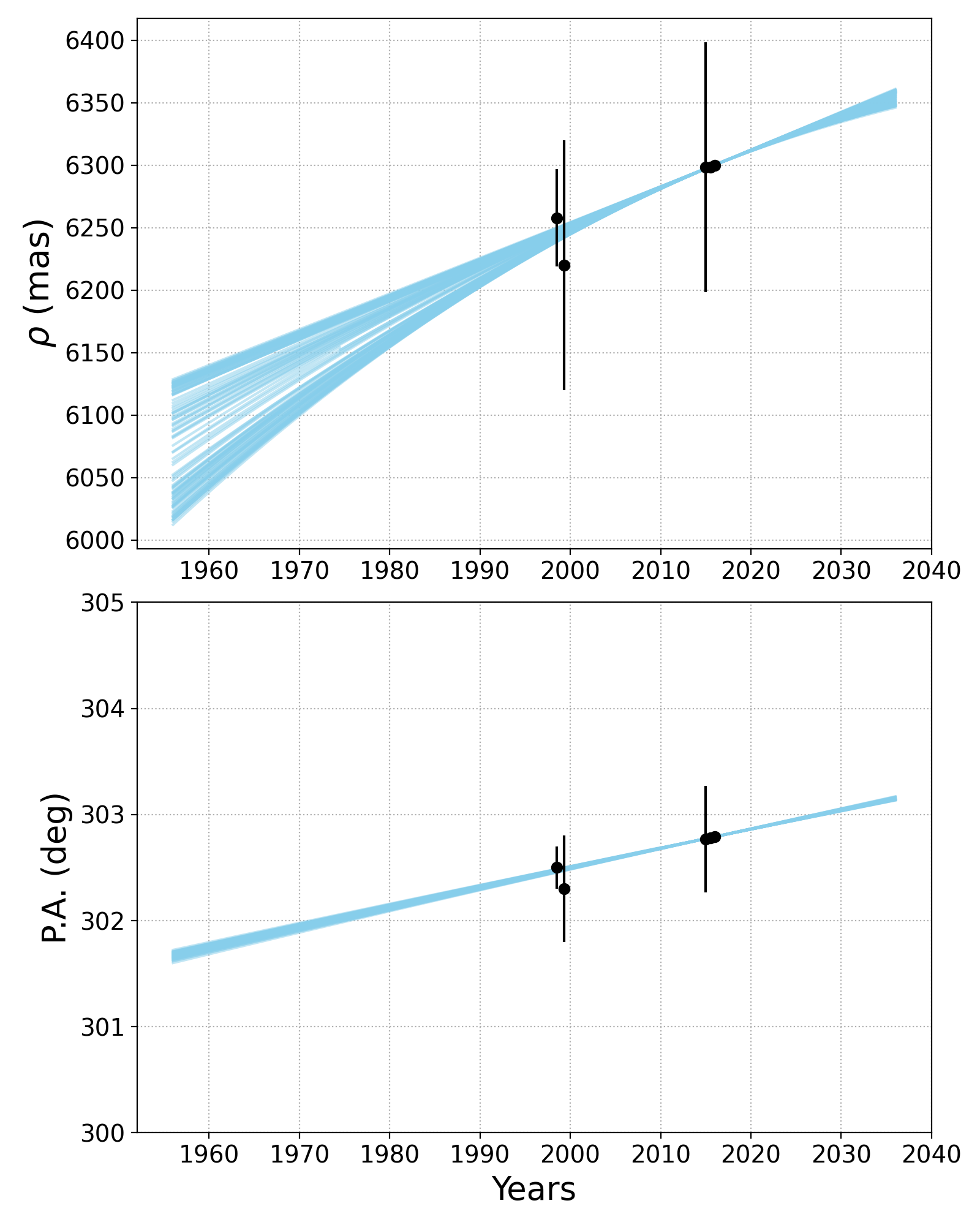}
	\label{AC_orbit_WDS}
    \end{minipage}
    \begin{minipage}[hbt!]{0.5\textwidth}
	\hspace{0.5cm}
	\includegraphics[width=.95\textwidth]{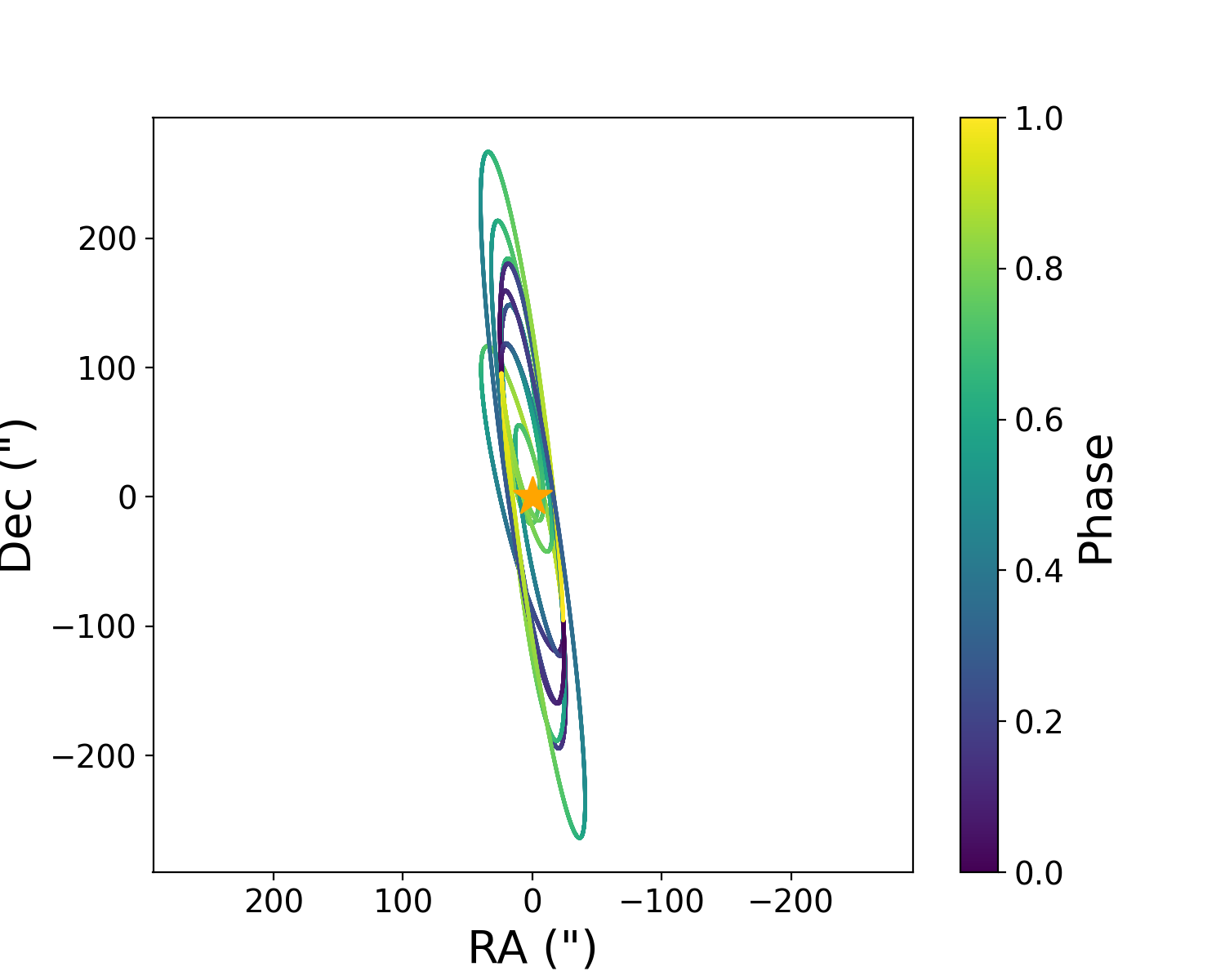}
	\label{AC_orbit}
	\includegraphics[width=.95\textwidth]{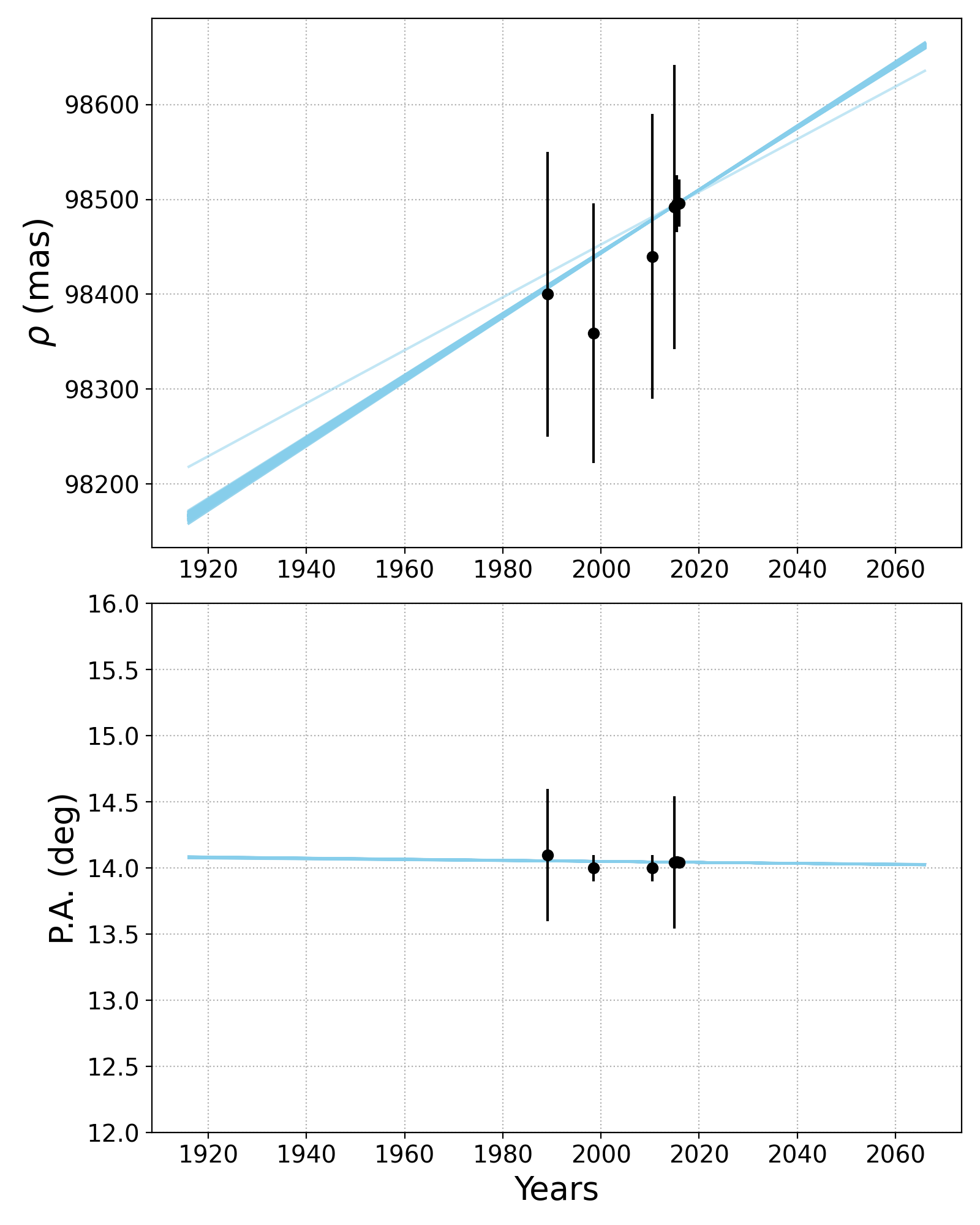}
	
	\label{AC_orbit_WDS}
    \end{minipage}
    \caption{{\textit{Top row:} AB (left) and AC (right) orbits with 10 random orbital solutions from posterior of the orbit fit. \textit{Middle row:} separation with WDS astrometry for AB (left) and AC (right) orbits with 100 random orbital solutions from posterior of the fit.\textit{Bottom row:} the same for the position angle.}} 
\end{figure*}

\begin{table*}
                \caption{Astrometry measurements from WDS \citep{WDS2001} for AC (left) and AB (right) systems used in the orbital fit.}
                \begin{minipage}{0.5\textwidth}
                \centering
                \begin{tabular}{lrcrc}
                \toprule
                \toprule
                        {Date} &  {sep}  & \bm{$\sigma_{\mathrm{sep}}$} & {P.A.}  & \bm{$\sigma_{\mathrm{PA}}$} \\
                        & $\bm{(\arcsec)}$ & $\bm{(\arcsec)}$ & $\bm {(^\circ)}$ & $\bm {(^\circ)}$\\
                          \toprule
                          \toprule
            1989.17 & 98.40 & 0.15 & 14.10 & 0.50 \\
            1998.52 & 98.36 & 0.14 & 14.00 & 0.10 \\
            2010.50 & 98.44 & 0.15 & 14.00 & 0.10 \\
            2015.00 & 98.49 & 0.15 & 14.04 & 0.50 \\
            2015.50 & 98.50 & 0.03 & 14.05 & 0.01 \\
            2016.00 & 98.50 & 0.03 & 14.04 & 0.01 \\
                \end{tabular} 
                \label{tab:appendix_ACastrometry}
                \end{minipage}
                \begin{minipage}{0.5\textwidth}
                \centering
                \vspace{-0.4cm}
                \begin{tabular}{lrcrc}
                \toprule
                \toprule
                        {Date} &  {sep}  & \bm{$\sigma_{\mathrm{sep}}$} & {P.A.}  & \bm{$\sigma_{\mathrm{PA}}$} \\
                        & $\bm{(\arcsec)}$ & $\bm{(\arcsec)}$ & $\bm {(^\circ)}$ & $\bm {(^\circ)}$\\
                          \toprule
                          \toprule
            1998.52 & 6.26 & 0.04 & 302.50 & 0.20 \\
            1999.29 & 6.22 & 0.10 & 302.30 & 0.50 \\
            2015.00 & 6.30 & 0.10 & 302.77 & 0.50 \\
            2015.50 & 6.30 & 1.20 10$^{-4}$ & 302.78 &1.00 10$^{-3}$ \\
            2016.00 & 6.30 & 3.30 10$^{-5}$ & 302.79 & 3.00 10$^{-4}$ \\
                \end{tabular} 
                \label{tab:appendix_ABastrometry}
                
                \end{minipage}
        \end{table*} 

\begin{table*}[hbt] 
\centering
\caption{Summary of orbital parameters for the TOI-4336 system. }
\label{table:orbital_solution}
\begin{tabular}{lccccccccc}
\toprule
\toprule
& \multicolumn{4}{c}{{AB system}} & \multicolumn{4}{c}{{AC system}}\\
\toprule
\toprule
\vspace{0.12cm}
\bf{Parameter}  & \bf{Median} & \bf{Std Dev} & \bf{68\% CI} & \bf{95\% CI} & \bf{Median} & \bf{Std Dev} & \bf{68\% CI} & \bf{95\% CI}\\
$a$ ($''$) & 5.9 & 29.9  &  (3.2, 7.8) & (3.2, 44.4) & 129.7 & 138.9 & (60.2, 160.1) & (59.6, 325.4)\\
$a$ (au) & 133 & 605  &  (72, 177) & (72, 1005) & 2915& 3120 & (1351, 3595) & (1340, 7309)\\
{Periastron (au)}  & {32.6} & {382.1}  &  {(5.1, 114.9)} & {(1.3, 434.1)} & {1705}& {1717} & {(169, 3765)} & {(25, 5493)}\\
{Period (years)} & {1908} & {33930} & {(752, 2886)} & {(746, 39282)} & {165150} & {418794} & {(52630, 227110)} & {(51972, 634155)}\\
$e$ & 0.79 & 0.26  &  (0.65, 0.98) & (0.12, 0.99) & 0.42 & 0.34 & (0.00005, 0.71) & (0.01, 0.99)\\
$i$ (deg) & 80.5 & 12.8 & (73.1, 85.5) & (45.8, 89.1) & 96.6 & 12.9 & (93.4, 100.1) & (92.3, 133.9)\\
$\omega$ (deg) & 182.8 & 101.4 & (82.5, 305.7) & (6.2, 342.9) & 178.7 & 107.8 & (146.3, 353.3) & (14.1, 359.9)\\
$\Omega$ (deg) & 123.1 & 112.2 & (-57.2, 137.9) & (-57.2, 299.8) & -165.5 & 104.9 & (-174.8, 9.0) & (-331.3, 14.0)\\
Distance (pc) & 22.44 & 0.02 & (22.42, 22.47) & (22.40, 22.49) & 22.46 & 0.02 &(22.44, 22.48) & (22.42, 22.49)
\vspace{0.12cm}
\end{tabular}
\end{table*}

\begin{figure*}[hbt]
    \centering
    \includegraphics[width=1.5\columnwidth]{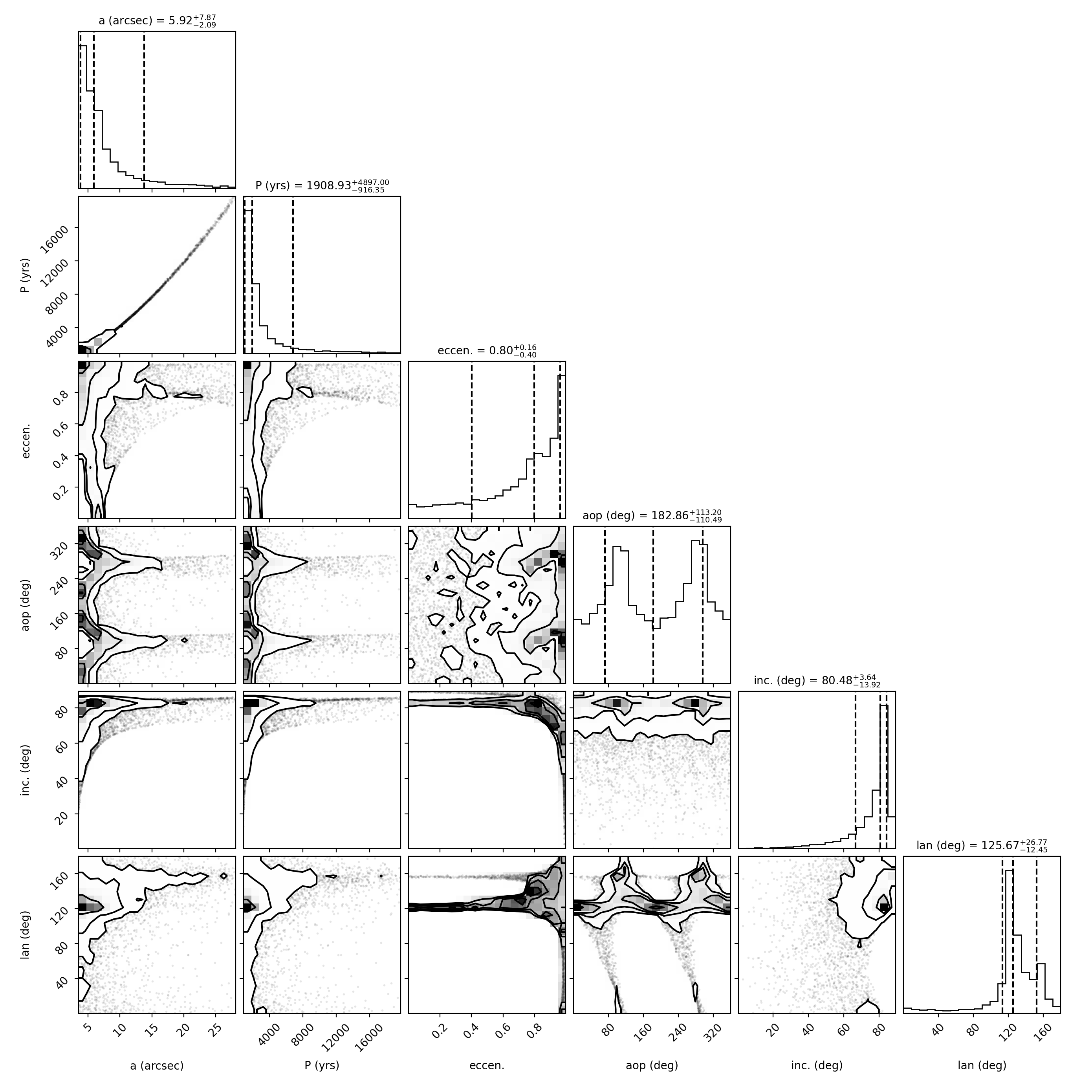}
    \caption{{Posterior samples of AB system. Histogram sub-panels show the posterior distribution, with the median and 68\% confidence intervals marked by dashed lines, with titles quantifying those ranges.}}
    \label{fig:corner_plot_AB}
\end{figure*}




\section{Global model errors and scaling factors}
\begin{table*}
\centering
\caption{{Global model errors and scaling factors for the MCMC analysis of the photometric data.}}
\label{table_baseline_error_fit}
\resizebox{\textwidth}{!}{
\begin{tabular}{lcccccc}
\toprule
\toprule
\vspace{0.15cm}
\bf{Observatory}  & \bf{Filter} & \bf{T$_0$ (BJD-TDB - 2 450 000)} & \bf{Baseline model} & \bf{Residual RMS (Exp. time)} & \bm{$\beta _w$} & \bm{$\beta _r$} \\ 
\toprule
\toprule
\vspace{0.12cm}
\tess Sector\,11  & TESS & 8616.7747 & p(t$^2$) & 1.21 10$^{-3}$ (120s) & 0.80 & {1.25} \\
\tess Sector\,38  & TESS & 9335.5733 & p(t$^2$) & 1.17 10$^{-3}$ (120s) & {0.99} & 1.01 \\
\tess Sector\,38  & TESS & 9351.9096 & p(t$^2$) & 1.26 10$^{-3}$ (120s) & 0.65 & {1.50} \\
TRAPPIST-South & \textit{Sloan-z'} & 9335.5733  &  p(t$^2$, a$^2$) & 2.59 10$^{-3}$ (15s) & 0.77 & 1.34 \\
SSO/Europa & \textit{Sloan-r'} & 9368.2460 & p(t$^2$,f$^2$) & 2.39 10$^{-3}$ (10s) & 0.76 & {1.37} \\
SSO/Ganymede & \textit{Sloan-r'} & 9368.2460 & p(t$^2$,f$^2$) & 2.63 10$^{-3}$ (10s) & {0.70} & {1.45} \\ 
TRAPPIST-South & \textit{Sloan-z'} & 9368.2460 & p(t$^2$,a$^2$) & 2.77 10$^{-3}$ (15s) & 0.72 & {1.40}\\
LCO (CTIO) & \textit{Pan-STARRS-zs} & 9368.2460 & p(t$^4$,a$^2$) & 1.41 10$^{-3}$ (45s) & {0.58} & {1.66} \\
TRAPPIST-South & \textit{Sloan-z'} & 9433.5913 & p(t$^2$,a$^2$) & 2.47 10$^{-4}$ (15s) & 0.90 & {1.13} \\
LCO (SAAO) & \textit{Sloan-g'} & 9629.9940 & p(t$^2$) & 2.38 10$^{-3}$ (150s) & 0.72 & 1.39 \\
SSO/Europa & \textit{Sloan-g'} & 9678.6362 & p(t$^2$) & 3.49 10$^{-4}$ (24s) & 0.69 & {1.46} \\
LCO (CTIO) & \textit{Sloan-g'} & 9678.6362 & p(t$^2$) & 1.70 10$^{-3}$ (150s) & 0.81 & {1.25} \\ 
LCO (CTIO)  & \textit{Pan-STARRS-zs} & 9678.6362 & p(t$^2$) & 1.33 10$^{-3}$ (45s) & 0.93 & {1.17}  \\
SSO/Europa & \textit{Sloan-r'} & 9727.6452 & p(t$^2$,f$^2$) & 2.40 10$^{-3}$ (45s) & 0.68 & {1.49}  \\
SSO/Io & \textit{Sloan-g'} & 9727.6452 & p(t$^2$) & 2.83 10$^{-3}$ (24s) & 0.77 & 1.35  \\
TRAPPIST-South & \textit{Sloan-z'} & 9727.6452 & p(t$^2$) & 1.92 10$^{-3}$ (15s) & 0.67 & {1.45} \\
ExTrA  & $1.2~\rm \mu m$ & 9727.6452 & p(t$^4$) & 2.29 10$^{-3}$ (60s) & {0.70} & {2.18} \\
ExTrA  & $1.2~\rm \mu m$ & 9727.6452 & p(t$^4$) & 1.94 10$^{-3}$ (60s) & {0.62} & {1.73} \\
SSO/Io & \textit{Sloan-r'} & 10021.6991 & p(t$^2$) & 3.32 10$^{-3}$ (10s) & 0.89 & 1.09\\
SSO/Europa & \textit{Sloan-r'} & 10021.6991 & p(t$^2$) & 3.24 10$^{-3}$ (10s) & 0.51 & {1.84}\\
LCO (CTIO) & \textit{Pan-STARRS-zs} & 10021.6991 & p(t$^4$) & 1.35 10$^{-3}$ (45s) & 0.72 & 1.38\\
LCO (CTIO) & \textit{Sloan-g'} & 10021.6991 & p(t$^2$) & 1.93 10$^{-3}$ (150s) & {0.54} & 1.87\\
    \tess Sector\,64  & TESS & 8616.7747 & p(t$^2$) & 1.33 10$^{-3}$ (120s) & 0.89 & 1.12 \\
\end{tabular}
}
\begin{flushleft}{For each transit observation, the chosen baseline model is denoted as p($\alpha^n$) with $\alpha$ a parameter such as $t$ = time, $a$ = airmass, $f$ = full width at half maximum of the point spread function, $x$ or $y$ = position of the target on the detector along the $x-$ and $y-$ axes, and $n$ the order of the considered polynomial. The residual RMS is also given with the corresponding exposure time and the error scaling factors $\beta_w$ and $\beta_r$. The column T$_0$ represents the expected mid-transit times.}\end{flushleft}
\end{table*}
\section{Posterior distributions of the transit model}
\begin{figure*}
    
    \centering
    \includegraphics[scale=0.25]{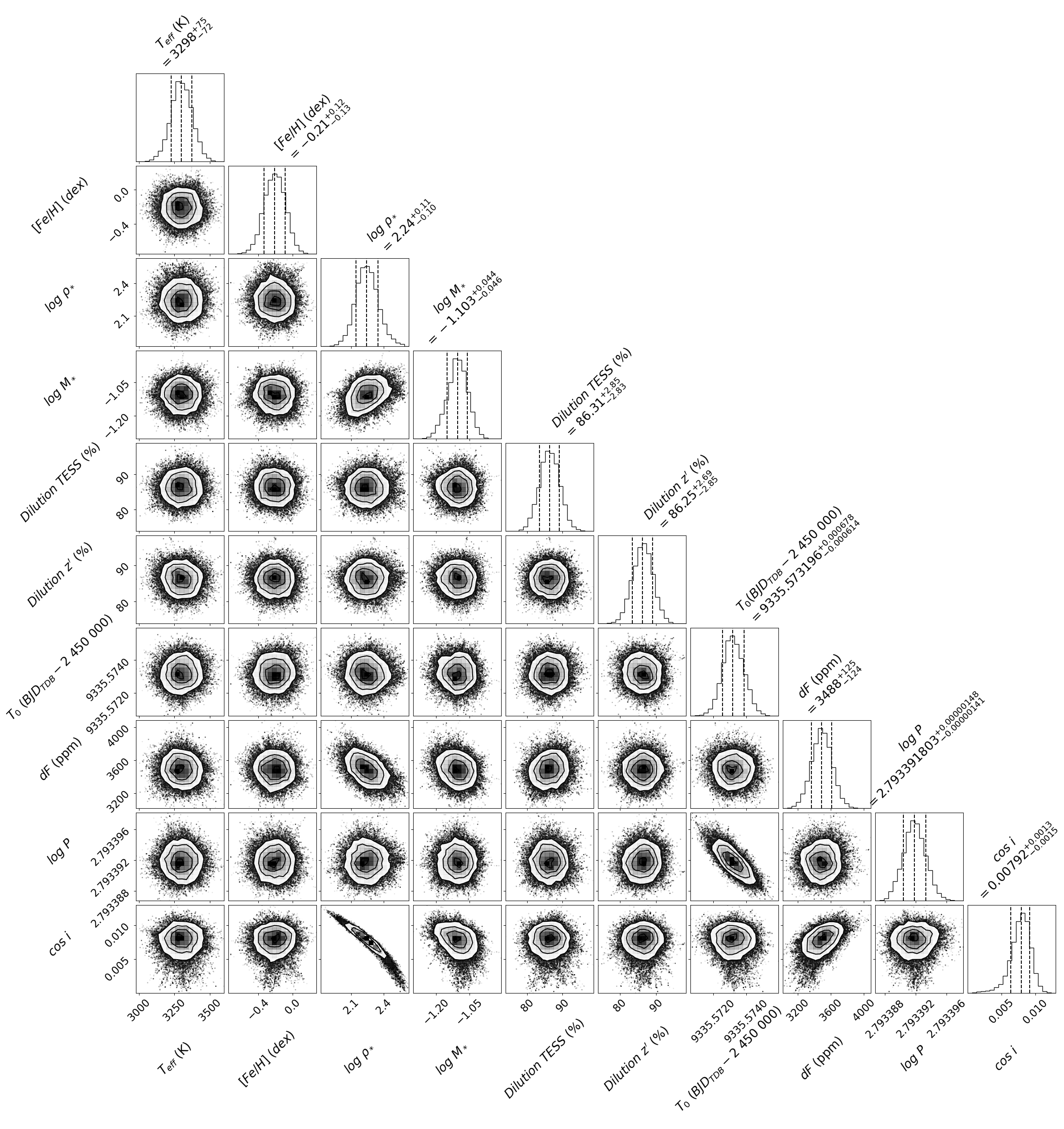}
    \caption{{Posterior distributions for the main jump parameters used in the \texttt{emcee} fit (see Section \ref{sec:global_analysis}): the effective temperature ($T_{\mathrm{eff}}$), the metallicity ([Fe/H]), the log of the stellar density (log $\rho_\star$), the log of the stellar mass (log $M_\star$), the dilution in the TESS and \textit{z'} bands, the epoch ($T_0$), the transit depth (dF), the log of the orbital period (log P) and the cosine of the orbital inclination of the planet (cos i).} }
     \label{fig:posteriors}  
\end{figure*}

\section{Transit Timing Variations}
\begin{figure*}
	\centering
	\includegraphics[scale=0.5]{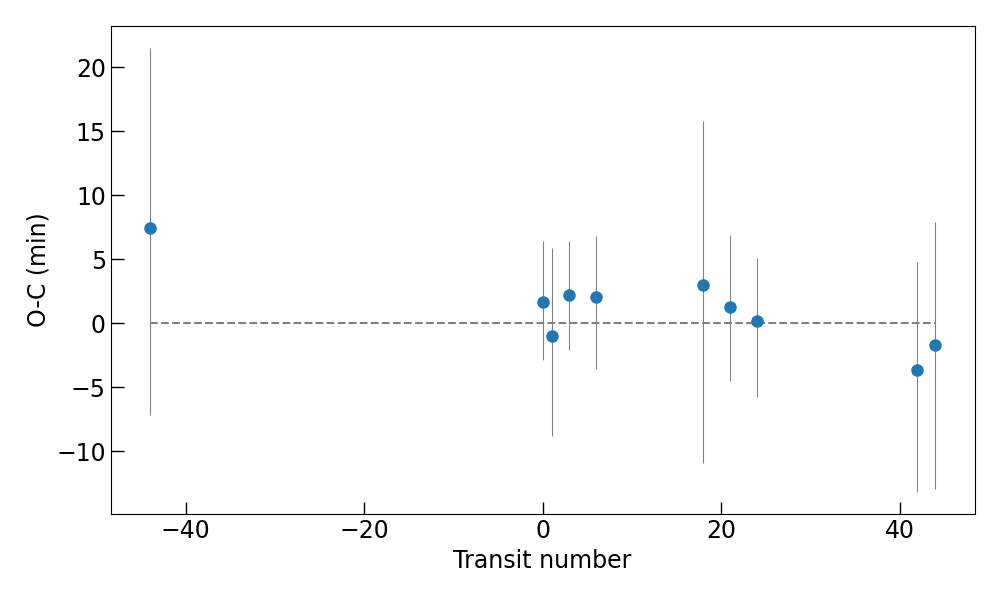}
	\caption{{O-C diagram obtained from a global fit allowing transit timing variations. We used as reference the timing obtained in the global transit model of Table \ref{tab:param}.} }
	\label{App:TTVs}
\end{figure*}

\section{Search for additional candidates in the system}
\begin{figure*}
    \centering
    \includegraphics[scale=0.55]{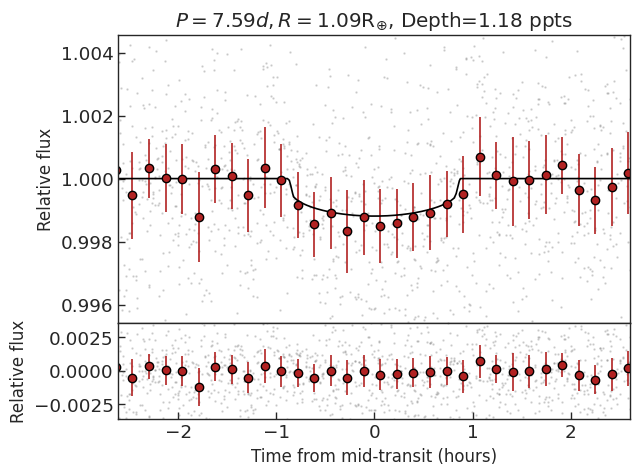}
    \caption{{{\textit{Top panel:}phase folded photometry of TIC 166184428.02. The best-fit model is shown in solid black. The grey points are the raw flux and the red points are the 15-minutes binned flux. \textit{Bottom panel:} residuals of the best-fit model.}}}
    \label{fig:allesfit_lc}
\end{figure*}

\begin{table*}
    \centering
    \caption{{Properties of TIC 166184428.02 obtained from the fitting stage of \sherlock, see Section \ref{ss:sherlock}. }}
        \begin{tabular}{lccc}
            \toprule
            \toprule
            {Parameter} & {Value} & {Unit} & {Source} \\ 
            \toprule
            \toprule
            \multicolumn{4}{c}{\textit{Fitted parameters}}\\ 
            \midrule
            $R_b / R_\star$ & $0.0306_{-0.0016}^{+0.0011}$ & & fit \\ 
            \vspace{0.1cm}
            $(R_\star + R_b) / a_b$ & $0.0310_{-0.0010}^{+0.0012}$ & & fit \\ 
            \vspace{0.1cm}
            $\cos{i_b}$ & $0.0069_{-0.0042}^{+0.0050}$ & & fit \\ 
            \vspace{0.1cm}
            $T_{0;b}$ & $2459333.2923\pm0.0025$ & $\mathrm{BJD}$& fit \\ 
            \vspace{0.1cm}
            $P_b$ & $7.587329_{-0.000041}^{+0.000030}$ & $\mathrm{d}$& fit \\ 
            \vspace{0.1cm}
            $q_{1; \mathrm{lc}}$ & $0.55_{-0.30}^{+0.27}$ & & fit \\ 
            \vspace{0.1cm}
            $q_{2; \mathrm{lc}}$ & $0.55_{-0.31}^{+0.27}$ & & fit \\ 
            \vspace{0.1cm}
            $\log{\sigma_\mathrm{lc}}$ & $-6.062\pm0.012$ & & fit \\ 
            \vspace{0.1cm}
            $\mathrm{gp: \ln{\sigma} (lc)}$ & $-7.14_{-0.21}^{+0.23}$ & & fit \\ 
            \vspace{0.1cm}
            $\mathrm{gp: \ln{\rho} (lc)}$ & $0.18\pm0.47$ & & fit \\ 
            \midrule
            \multicolumn{4}{c}{\textit{Derived parameters}} \\ 
            \midrule 
            Host radius over semi-major axis; $R_\star/a$ & $0.0301_{-0.0010}^{+0.0012}$ & & derived \\ 
            \vspace{0.1cm}
            Semi-major axis over host radius; $a/R_\star$ & $33.2_{-1.3}^{+1.2}$ & & derived \\ 
            \vspace{0.1cm}
            Companion radius over semi-major axis; $R_p/a$ & $0.000920_{-0.000052}^{+0.000047}$ & & derived \\ 
            \vspace{0.1cm}
            Companion radius; $R_p$  & $1.088_{-0.064}^{+0.051}$ & $\mathrm{R_{\oplus}}$ & derived \\ 
            \vspace{0.1cm}
            Semi-major axis; $a$ & $0.0504\pm0.0024$ & au & derived \\ 
            \vspace{0.1cm}
            Inclination; $i$  & $89.60_{-0.29}^{+0.24}$ & deg & derived \\ 
            \vspace{0.1cm}
            Impact parameter; $b$ & $0.23_{-0.14}^{+0.16}$ & & derived \\ 
            \vspace{0.1cm}
             Total transit duration; $T_\mathrm{tot}$ & $1.739_{-0.083}^{+0.075}$ & h & derived \\ 
             \vspace{0.1cm}
             Full-transit duration; $T_\mathrm{full}$  & $1.631_{-0.087}^{+0.074}$ & h & derived \\ 
             \vspace{0.1cm}
             Host density from orbit; $\rho_\mathrm{\star}$ & $12.0\pm1.3$ & cgs & derived \\ 
             \vspace{0.1cm}
             Equilibrium temperature; $T_\mathrm{eq}$ & $378\pm24$ & K & derived \\ 
             \vspace{0.1cm}
             Transit depth (undil.); $\delta_\mathrm{tr; undil}$ & 
             $1.18_{-0.12}^{+0.14}$ & ppt & derived \\ 
             \vspace{0.1cm}
             Limb darkening; $u_\mathrm{1; lc}$ & $0.73_{-0.42}^{+0.47}$ &  & derived \\ 
             \vspace{0.1cm}
             Limb darkening; $u_\mathrm{2; lc}$ & $-0.07_{-0.37}^{+0.43}$ & & derived \\ 
        \end{tabular}
        \label{tab:allesfit.02}
\end{table*} 

\begin{figure*}
    \centering
    \includegraphics[scale=0.2]{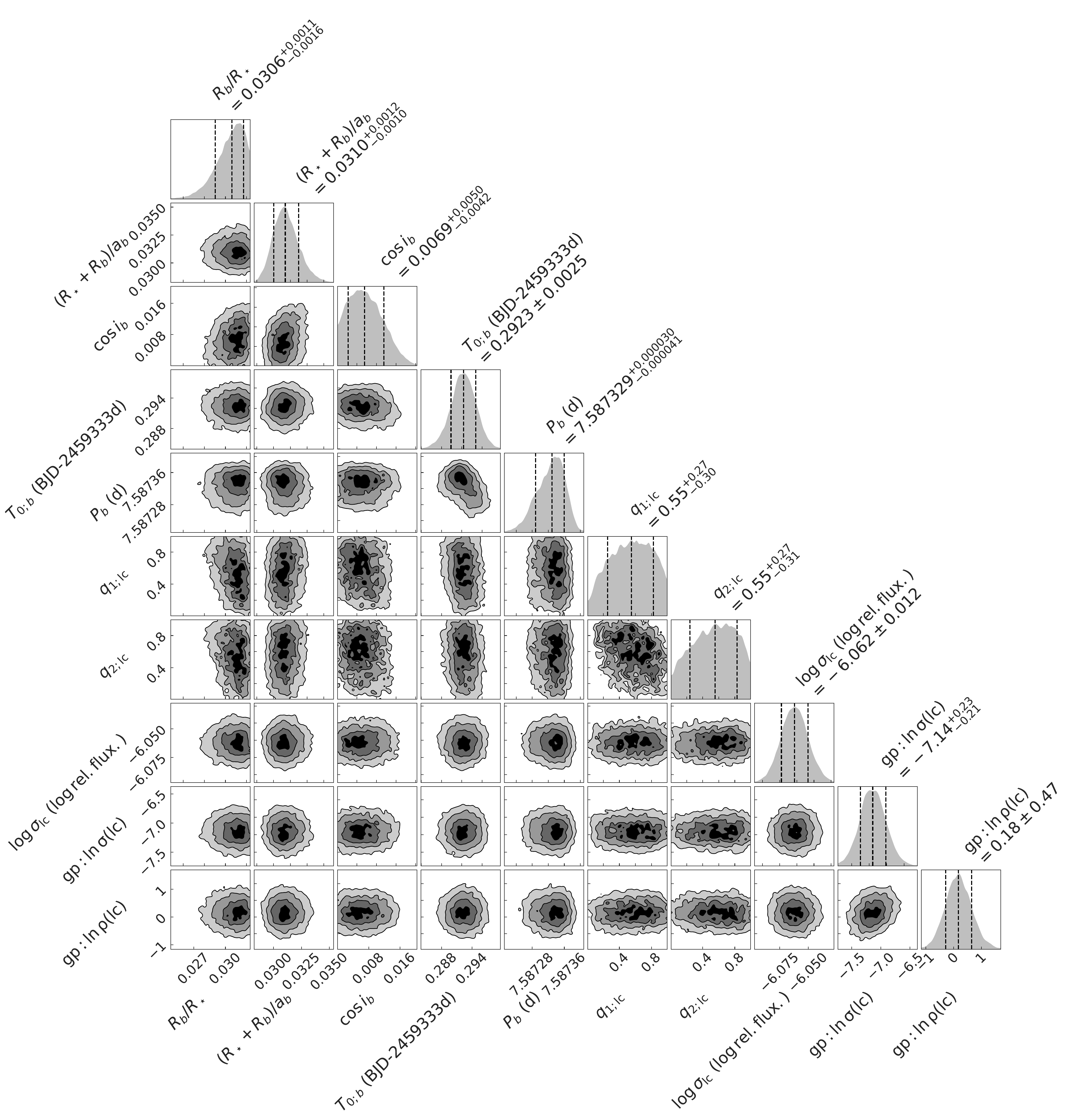}
    \caption{{Posterior distributions for the main parameters fitted in the Nested Sampling fit for the second candidate found in the TOI-4336 system, TIC 166184428.02 (see Section \ref{ss:sherlock}), with $R_b$ the candidate radius, $R_\star$ the stellar radius, $i_b$ the inclination, $T_{0; b}$ the epoch, $P_b$ the period, $q_{1;lc}$ and $q_{2;lc}$ the Kipping parametrisation \citep{kippingldcs} for the limb darkening coefficients, $log \sigma_lc$ the error scaling parameter, and $ln \sigma(lc)$ and $ln \rho(lc)$ the two GP parameters related to the amplitude and length scale.}}
    \label{fig:allesfit_corner}
\end{figure*}

\end{appendix}
%
%






   
  



\end{document}